\documentclass[AMA,STIX1COL]{WileyNJD-v2}
\usepackage{moreverb}
\usepackage{amsfonts}
\usepackage{fontenc}
\usepackage[intlimits]{amsmath}
\usepackage[utf8]{inputenc}
\usepackage{graphicx}
\usepackage{enumerate}
\usepackage{multirow}
\usepackage{multicol}
\usepackage{xcolor}
\usepackage{tikz}
\usepackage{url} % not crucial - just used below for the URL 
\usepackage[figureposition=bottom,tableposition=top]{caption}
\usepackage{subcaption}
\usepackage{mathtools}
\usepackage{array}
\usepackage{xurl}

\newcommand{\btheta}{ \mbox{\boldmath $\theta$}}

\newcommand{\bc}{ \mbox{\bf c}}

\newcommand{\by}{ \mbox{\bf y}}

\newcommand{\bs}{ \mbox{\bf s}}

\newcommand{\bu}{ \mbox{\bf u}}

\newcommand{\iid}{\stackrel{iid}{\sim}}

\newcommand{\beq}{ \begin{equation}}
\newcommand{\eeq}{ \end{equation}}
\newcommand{\beqn}{ \begin{eqnarray}}
\newcommand{\eeqn}{ \end{eqnarray}}

\newcommand{\rtwo}{\mbox{${\mathbb{R}^{2}}$}}

\newcommand{\rpm}{\sbox0{$1$}\sbox2{$\scriptstyle\pm$}
  \raise\dimexpr(\ht0-\ht2)/2\relax\box2 }

\def\Lp{\left(}
\def\Rp{\right)}

\newcommand\BibTeX{{\rmfamily B\kern-.05em \textsc{i\kern-.025em b}\kern-.08em
T\kern-.1667em\lower.7ex\hbox{E}\kern-.125emX}}

\articletype{Article Type}%

\received{<day> <Month>, <year>}
\revised{<day> <Month>, <year>}
\accepted{<day> <Month>, <year>}

\begin{document}

\title{Multivariate cluster point process to quantify and explore multi-entity configurations: Application to biofilm image data}

\author[1]{Suman Majumder*}

\author[2]{Brent A. Coull}

\author[3]{Jessica L. Mark Welch}

\author[4]{Patrick J. La Riviere}

\author[3]{Floyd E. Dewhirst}

\author[5]{Jacqueline R. Starr$^{0,}$}

\author[2]{Kyu Ha Lee$^{0,}$}

\authormark{Majumder \textsc{et al}}

\address[1]{University of Missouri, Missouri, USA}

\address[2]{Harvard T.H. Chan School of Public Health, Massachusetts, USA}

\address[3]{Forsyth Institute, Massachusetts, USA}

\address[4]{University of Chicago, Illinois, USA}

\address[5]{Brigham and Women's Hospital, Massachusetts, USA}

\corres{*Suman Majumder, University of Missouri. \email{sm8qr@missouri.edu}}

%\presentaddress{Present address}

\abstract[Abstract]{Clusters of similar or dissimilar objects are encountered in many fields. Frequently used approaches treat each cluster’s central object as latent. Yet, often objects of one or more types cluster around objects of another type. Such arrangements are common in biomedical images of cells, in which nearby cell types likely interact. Quantifying spatial relationships may elucidate biological mechanisms. Parent-offspring statistical frameworks can be usefully applied even when central objects (“parents”) differ from peripheral ones (“offspring”). We propose the novel multivariate cluster point process (MCPP) to quantify multi-object (e.g., multi-cellular) arrangements. Unlike commonly used approaches, the MCPP exploits locations of the central parent object in clusters. It accounts for possibly multilayered, multivariate clustering. The model formulation requires specification of which object types function as cluster centers and which reside peripherally. If such information is unknown, the relative roles of object types may be explored by comparing fit of different models via the deviance information criterion (DIC). In simulated data, we compared a series of models’ DIC; the MCPP correctly identified simulated relationships. It also produced more accurate and precise parameter estimates than the classical univariate Neyman-Scott process model. We also used the MCPP to quantify proposed configurations and explore new ones in human dental plaque biofilm image data. MCPP models quantified simultaneous clustering of \emph{Streptococcus} and \emph{Porphyromonas} around \emph{Corynebacterium} and of \emph{Pasteurellaceae} around \emph{Streptococcus} and successfully captured hypothesized structures for all taxa. Further exploration suggested the presence of clustering between \emph{Fusobacterium} and \emph{Leptotrichia}, a previously unreported relationship.}


\jnlcitation{\cname{%
\author{S. Majumder}, 
\author{B. A. Coull}, 
\author{J. L. Mark Welch}, 
\author{P. J. La Riviere},
\author{F. E. Dewhirst},
\author{J.R. Starr},and 
\author{K. Lee}} (\cyear{2024}), 
\ctitle{Multivariate cluster point process to quantify and explore multi-entity configurations: Application to biofilm image data}, \cjournal{Stat. In Medicine}, \cvol{2024;00:1--6}.}

\footnotetext{Co-senior authors}

\maketitle

\section{Introduction}
\label{s: intro}

Instances abound in nature where one type of object is dispersed around another type of object. At interplanetary and field ecologic scales, respectively, radio galaxies \citep{yates1989cluster,hill1991change} and forest songbirds \citep{tarof2004habitat} exemplify spatial clustering. At microscopic scales, among human cells and single-celled organisms, such arrangements are common. For example, hair cell development in the inner ear depends on surrounding support cells \citep{fritzsch2010canal}, neurons generally require various adjacent support cells \citep{molnar2015concepts}, granulomatous lesions of the lung exhibit concentric rings of different cell types \citep{gideon2019neutrophils}, and platelets aggregate around red blood cells during clot formation, which can also involve other cell types. To describe and quantify such arrangements, we have developed a new multivariate cluster point process model (MCPP). We also developed a procedure for using the MCPP to  explore possible clustering relationships not known \emph{a priori}. In human dental plaque biofilm, spherical \textit{Streptococcus} cells often cluster  around the ends of filamentous \textit{Corynebacterium} cells \citep{jones1972special, welch2016biogeography}. These corncob-like arrangements have been observed for decades and likely hold clues about microbial interactions that can affect human health. We assessed performance of the MCPP through simulations and by application to dental plaque biofilm image data.

Most biofilm image analysis approaches have focused on  macro-level structural characteristics or derived features, such as biofilm volume, thickness, or surface roughness \citep{vorregaard2008comstat2,hartmann2021quantitative}. Less commonly, spatial point process models have been used to analyze the spatial patterning of microbial cells: how cells of one taxonomic class (or taxon) are distributed in relation to other cells of the same or different taxa. However, standard point process models, such as the log-Gaussian Cox process model \citep{moller1998log}, do not account for the complex spatial clustering arrangements often present in biofilm images. Methods that can account for such complexity are needed to quantify visible arrangements and to explore and quantify spatially dependent arrangements that, unlike the ``corncobs", are not visually discernable. Quantification, in turn, would allow benchmarking and hypothesis testing, such as for comparing effects of an antibacterial treatment on biofilm architecture.

The Neyman-Scott process model (NSP) \citep{neyman1958statistical} is a classic statistical model used to quantify spatially dependent clustering relationships, most often applied to actual parents and offspring (such as trees in a forest). In this approach, locations of the central object in each cluster are treated as latent, in part because in a cluster comprising similar objects (such as a grove of trees or a school of fish), it may be impossible to identify any true parent individuals.  In contrast, in corncob arrangements in dental plaque biofilm, the locations of the central cells are often known. Though reasonable for within-taxon clustering, na\"ive application of the NSP model to investigate between-taxon relationships is inappropriate because it ignores the taxon in the center of the clusters.

Corncob arrangements in dental plaque biofilm have other characteristics that preclude direct application of the NSP model or the related shot noise Cox process model \citep{moller2003shot}. First, the “parent” (i.e., central) and “offspring” (i.e., peripheral) cells are of different bacterial taxa, thus requiring a multivariate extension of univariate approaches. Though others have considered point process models for multi-type spatial data \citep{berman1986testing, moller1998log, grabarnik2009modelling,diggle2013spatial}, we are unaware of any work addressing more complex clustering arrangements as a specific focus of model development. Second, the corncob arrangements sometimes include multiple “offspring” taxa in that both \emph{Streptococcus} and \emph{Porphyromonas} are observed around \emph{Corynebacterium} “parents.” Existing multivariate cluster point process models \citep{tanaka2014identification, jalilian2015multivariate} can not model the taxon in the cluster center and have no scope to enforce multiple offspring taxa's having the same parent taxon. 

\begin{figure}
    \centering
    \includegraphics[width=0.3\linewidth]{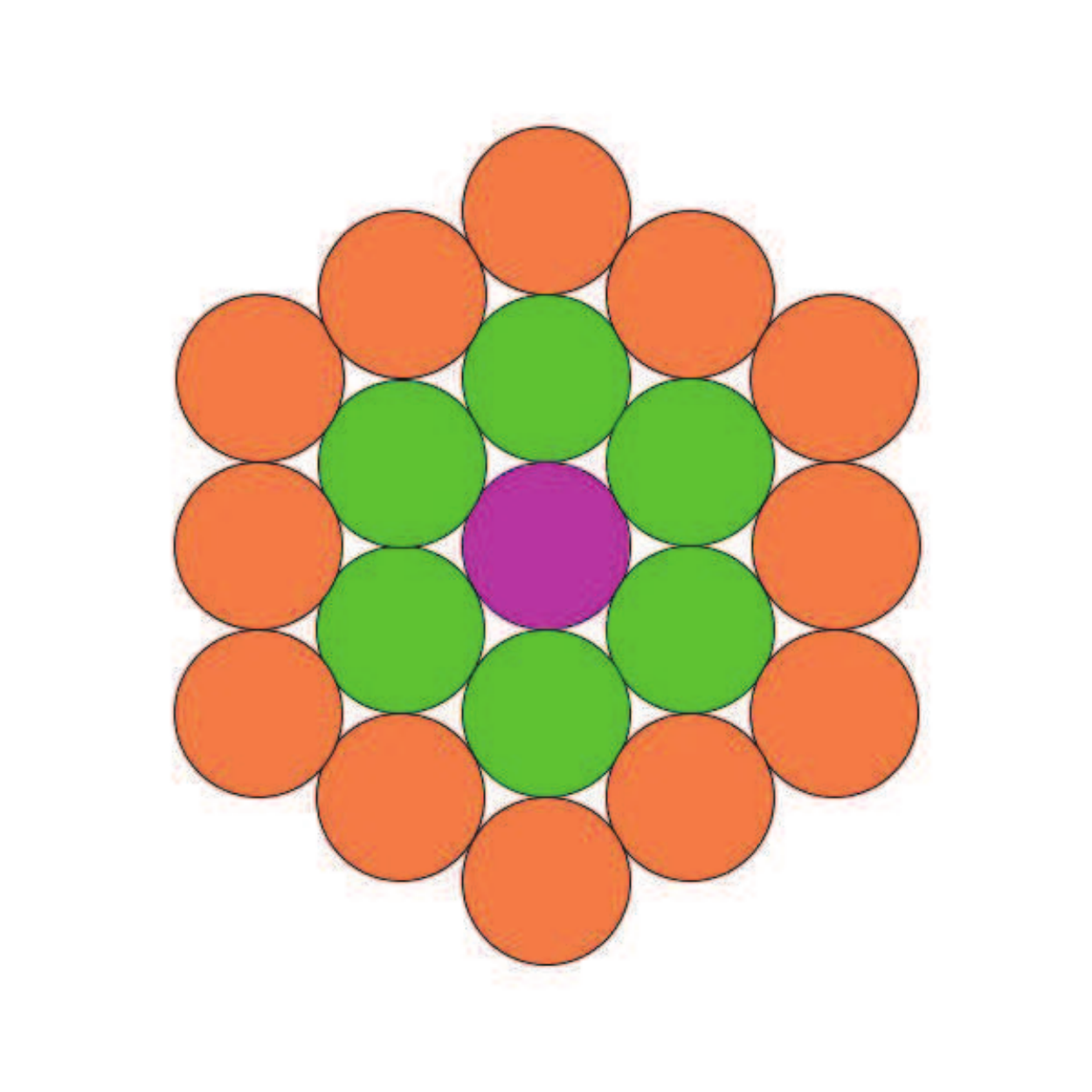}
    \caption{A multi-layered parent-offspring structure. The green dots behave as offspring to the pink dot while the orange dots behave as offspring to the green dots}
    \label{fig:multilayer_diag}
\end{figure}

Illian et al. \cite{illian2009hierarchical} propose a multivariate clustering process model that permits multiple types of offspring to be clustered around the center. In the proposed configuration, however, any offspring can be affected by all potential cluster centers. Thus these models fail to exploit biological knowledge about which taxa are parents to which other offspring taxa. Further, parent processes are considered as fixed, known quantities and are not explicitly modeled. To our knowledge, existing approaches fail to address a third challenge, that the corncobs can be multilayered in nature (e.g., Figure \ref{fig:multilayer_diag} displays a synthetic toy example of such configuration). Fourth, the corncob arrangements themselves are part of a more complexly organized community that includes other taxa unrelated to the arrangements. None of the aforementioned methods address this challenge. Fifth, clustering configurations are not always known \emph{a priori}. When configurations are known, the analysis goal might be to quantify them; in other instances the goal might be to explore the possible existence of clustering relationships.

The primary innovation of the proposed MCPP model is that it exploits the known locations of central objects in clusters. It also addresses the above-described challenges together. Specifically, it can simultaneously quantify multivariate and multilayered clustering and inter-process relationships. The MCPP is flexible in that it can simultaneously model clustered and non-clustered processes and can be applied in multivariate (multi-taxon) or univariate (individual taxon) contexts. It may seem restrictive that fitting the MCPP model requires specification of which objects are central to clusters and which peripherally located. We demonstrate that the MCPP can also be used to explore and infer  the presence of newly proposed relationships by comparing model fit via the deviance information criterion (DIC). In this paper, we describe  the model in detail, evaluate its performance through simulation studies, and demonstrate the feasibility of applying the model to both real and synthetic datasets.

%\vspace{-1.5 em}

%\vspace{-1.75 em}
\begin{figure}
    \centering
    \includegraphics[width=\linewidth]{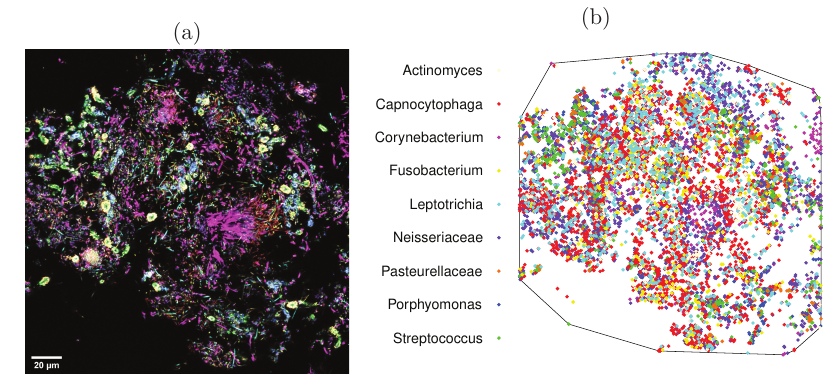}
    % \begin{subfigure}[b]{0.4\linewidth}
    % \caption{}
    % \includegraphics[width=\linewidth]{pics/alldat_RGB.eps}
    % \end{subfigure}
    % %\hfill
    % \begin{subfigure}[b]{0.59\linewidth}
    % \caption{}
    % \includegraphics[width=\linewidth,trim={0.5cm 0.5cm 0.5cm 0.5cm},clip]{pics/alldat_noquad.eps}
    % \end{subfigure}
    % \flushright \begin{subfigure}[b]{0.59\linewidth}
    % \includegraphics[width=\linewidth]{pics/Rout_noEub.jpeg}
    % \caption{}
    % \end{subfigure}
    \caption{(a) A biofilm image of a dental plaque sample from from a human donor: RGB image - taxa relevant to the multilayered concob arrangements are \textit{Corynebacterium} (Pink), \textit{Streptococcus} (Green), \textit{Porphyromonas} (Blue) and \textit{Pasteurellaceae} (Orange); (b) post-segmentation spatial locations of the centroids of cells identified via genus-specific probes, as indicated. Convex hull of all the centroids form the boundaries of the analysis window.}
    \label{fig:G1_comp}
\end{figure}
%\vspace{-3 em}
\section{Microbiome Biofilm Image Data from Dental Plaque Samples}
\label{s: data_des}
\begin{figure}
    \centering
    \includegraphics[width=\linewidth]{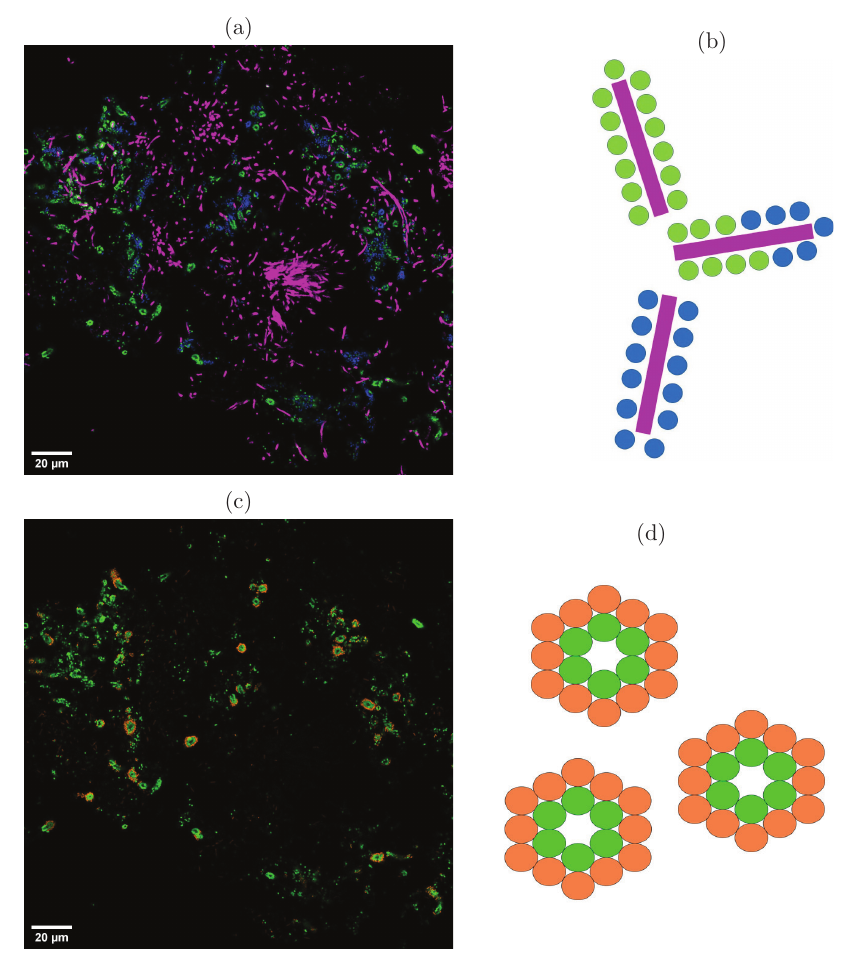}
    % \begin{subfigure}[h]{0.53\linewidth}
    % \caption{}
    % \includegraphics[width=\linewidth]{pics/CSP_all.eps}
    % \end{subfigure} 
    % %\hspace{5.5 em}
    % \hfill
    % \begin{subfigure}[h]{0.3\linewidth}
    % \caption{}
    % \includegraphics[width=\linewidth]{pics/corncobs_cartoon.eps}
    % %\vspace{2 em}
    % \end{subfigure}
    % \begin{subfigure}[h]{0.53\linewidth}
    % \caption{}
    % \includegraphics[width=\linewidth]{pics/SP.eps}
    % \end{subfigure}
    %  \hfill
    %  \begin{subfigure}[h]{0.45\linewidth}
    %  \caption{}
    %  \includegraphics[width=\linewidth]{pics/sp_woc.eps}
    %  \end{subfigure}

    \caption{(a) The same image as in Figure \ref{fig:G1_comp}, displaying channels corresponding to \textit{Corynebacterium} (Pink), \textit{Streptococcus} (green) and \textit{Porphyromonas} (Blue); (b) cartoon depicting various corncob structures visible in the image. The corncob structures are present in areas of the image corresponding to the tip of the \textit{Corynebacterium}. At the center of the dense clump of \emph{Corynebacterium} in (a), no corncobs are observed; (c) The same image as in Figure \ref{fig:G1_comp}, displaying channels corresponding to \textit{Pasteurellaceae} (Orange)  and \textit{Streptococcus} (Green); and  (d) a cartoon image illustrating the clustering of \textit{Pasteurellaceae} around \textit{Streptococcus}. Hiding the pink \textit{Corynebacterium} channels highlights the outer layer in the multilayer arrangements. For other examples of biofilm images with corncob structures in which both Streptococcus and \emph{Porphyromonas} cells cluster around the same \emph{Corynebacterium} cell, see Supplemental Figure S2, panel C, in Mark Welch et al. (2016).}
    \label{fig:G1_CSP_SPo}
\end{figure}

In this section, we present image data that motivate development of the method and that we analyzed by applying the MCPP. These data and their collection methods and processing have been described in detail elsewhere \citep{welch2016biogeography}. Briefly, dental plaque samples were collected, embedded in methacrylate, sectioned, and subjected to multi-spectral fluorescence \textit{in situ} hybridization (FISH) imaging. Multiple genus- and family-level probes were applied simultaneously to identify nine taxa. The near-universal probe Eub338 targeting most bacteria was also included (Figure \ref{fig:G1_comp}(a)). We created the scale and the subsequent grid for the image by setting 9.64 pixels to be equal 1 to $\mu m$. The bottom, left corner of the image is assigned the coordinate (0,0), and the grid is formed using the $\mu m$ scale. This scale is maintained in Figures \ref{fig:G1_comp}(a)-(b), \ref{fig:G1_CSP_SPo}(a) and (c) as well as Supplementary Figures S.1, S.2 and S.9. We performed segmentation of the biofilm image in FIJI \citep{schindelin2012fiji} by applying a $3\times 3$ median filter and the ``Auto Local" thresholding function with the Bernsen method (See the ImageJ website for details: \url{https://imagej.net/plugins/auto-local-threshold}). Spatial coordinate information for each cell's centroid was generated by applying the ``Analyze Particles" function with size filter of 0.5 $\mu m$ diameter in FIJI (Figure \ref{fig:G1_comp}(b)). The window $\mathcal{W}$, required for the statistical analysis, is created by taking a convex hull of the observed centroid locations without altering the grid or the coordinates (Figure \ref{fig:G1_comp}(b)). 

The sampled image is representative of similar samples from the same and other donors without active tooth decay (not shown). Filamentous \textit{Corynebacterium} cells (Figure \ref{fig:G1_CSP_SPo}, pink) are common. Members of this genus are now believed to help establish the ``healthy" dental microbiota, providing a scaffold around which other community members assemble. Some \textit{Streptococcus} (Figure \ref{fig:G1_CSP_SPo}, green) and/or \textit{Porphyromonas} (Figure \ref{fig:G1_CSP_SPo}, blue) cells surround the tips of \textit{Corynebacterium} filaments \citep{welch2016biogeography}. Adding another layer and further complexity, \textit{Pasteurellaceae} cells (Figure \ref{fig:G1_CSP_SPo}, orange) sometimes surround \textit{Streptococcus} cells \citep{welch2016biogeography,morillo2022corncob}. These structures are observed in the real data (Figures \ref{fig:G1_CSP_SPo} (a) and (c)) and are illustrated with cartoons in Figures \ref{fig:G1_CSP_SPo} (b) and (d). The side-by-side images are not intended to be at the same scale. As outlined in Section \ref{s: intro}, these arrangements allow \textit{Streptococcus} to serve as either parent, offspring, or both in the same cluster. Other taxa scatter seemingly homogeneously and may have additional spatial relationships with \textit{Corynebacterium}, \textit{Streptococcus}, \textit{Porphyromonas}, or \textit{Pasteurellaceae} (Supplementary Materials, Section B). Below, we demonstrate a procedure to explore previously undescribed patterns among these other taxa.
%\vspace{-1 em}

%\vspace{-2.75 em}
\section{Multivariate  Cluster  Point  Process Model}\label{s: method}
We consider a multivariate process $Y$ to be a collection of processes $Y_i \,,\ i=1, \ldots, m$, at location $\bs \in \mathcal{W} \subset \rtwo$, where each component $Y_i$ is a Poisson process characterized by an intensity function $\lambda_i(\bs)$ and $\mathcal{W}$, the observation window. In application to the biofilm sample, each $Y_i$ captures the spatial distribution of the $i$-th taxon.  

%\vspace{-1 em}
\subsection{Model formulations} \label{s: model}
Suppose that $Y_1, \ldots, Y_p$, are homogeneous Poisson processes (HPP) with intensities $\lambda_v^C$ for $v=1, \ldots, p~(<m)$, and they serve only as parent processes (``C" signifies central objects, or parents). Then we consider the $q \,\  (\leq m-p)$ processes, $Y_{p+1}, \ldots, Y_{p+q}$, that behave as offspring processes. Each offspring process is assumed to have one parent process. Let $C_l$ for $l=p+1, \ldots, p+q$ denote the corresponding parent process for offspring process $Y_{l}$ with the following properties:
%\vspace{-1 em}
\begin{enumerate}[i)]
    \item If two offspring processes $Y_l$ and $Y_{l'}$ , $l \neq l'$ share a parent process, then we label their parent processes to be same, i.e., $Y_l = Y_{l'}$;
    \item If the $j$-th process $Y_j$ serves as a parent process for the $l$-th process $Y_l$ for some $j$ and $l \in {p+1, ... , p+q}$, then we label the $l$-th parent process $C_l$ to be same as the $j$-th process $Y_j$ , i.e., $C_l = Y_j$ .
\end{enumerate}
%\vspace{-1.5 em}
We assume that for each parent of the $l$-th offspring process $Y_l$ located at $\bc_l \in C_l \,,\ l \in \{p+1, \ldots, p+q\}$, its offspring are distributed around it according to formula $\alpha_lk_l(\cdot - \bc, h_l)$, where $\alpha_l$ is the average number of offspring per parent, $k_l( \cdot, \cdot)$ is a kernel (e.g. Gaussian, uniform or Cauchy), and $h_l$ is a bandwidth parameter that controls the distance between the parent and its offspring locations for the $l$-th offspring process \citep{chiu2013stochastic}. The offspring process $Y_l$ is therefore the union of all such offsprings. The remaining $m-p-q$ types (e.g. taxa) that are unrelated to multilayered arrangements are modeled as HPP with intensities $\lambda_j$ for $j = p+q+1, \ldots, m$.  Therefore, under the proposed specification, the intensity functions for the various processes are
%\vspace{-0.5 em}
\beq \label{eq: all_taxa}\begin{split}
    \lambda_v(\bs) &= \lambda_v^C \,,\ v = 1, \ldots, p \\
    \lambda_l(\bs) &= \alpha_l \sum_{\bc_l \in C_l} k_l(\bs - \bc_l,h_l) \,,\ l = p+1, \ldots, p+q,\\
     \lambda_j(\bs) &= \lambda_j \,,\ j = p+q+1, \ldots, m.
\end{split}
\eeq

Since the different processes are assumed to be conditionally independent of each other, the likelihood as a function of the unknown parameters, $\btheta$ = $\Big\{\alpha_{p+1}$, $\ldots$ ,$\alpha_{p+q}$,$h_{p+1}$, $\ldots$ , $h_{p+q}$, $\lambda_1^C$, $\ldots$ ,$\lambda_p^C$, $\lambda_{p+q+1}$, $\ldots$ ,$\lambda_{m}\Big\}$, is given as the product of the individual likelihoods for each of the processes: \beq \label{eq: Lnolog}
L(Y|\btheta) = \prod_{i=1}^m \left[\int_{\mathcal{W}} (1-\lambda_i(\bu)) \,\ d\bu \left(\prod_{\bs \in Y_i} \lambda_i(\bs)\right)\right].
\eeq Using the intensities $\lambda_i(\bs)$ defined based on (\ref{eq: all_taxa}), and taking logarithm simplifies the log-likelihood to be \beq \label{eq: L}
\begin{split}
    l(Y|\btheta) &\propto | \mathcal{W} | -\sum_{v=1}^{p} | \mathcal{W} |\lambda^C_v - \sum_{l=p+1}^{p+q} \alpha_l\sum_{\bc_l \in C_l}\int_{\mathcal{W}}k_l(\bu - \bc_l,h_l) \text{d}\bu - \sum_{j=p+q+1}^m | \mathcal{W} | \lambda_j\\
     & + \sum_{v=1}^{p} n_v\log \lambda^C_v + \sum_{l=p+1}^{p+q} \sum_{\by \in Y_l} \log \Lp \alpha_l\sum_{\bc_l \in C_l} k_l(\by - \bc_l,h_l) \Rp +  \sum_{j=p+q+1}^m n_j\log \lambda_j,
\end{split}
\eeq  
where $| \mathcal{W} |$ is the area of $\mathcal{W}$, and $n_i$ denotes the number of observations from the $i$-th process in $\mathcal{W}$.

\subsection{Quantities of interest} \label{s: interpret}

For a stationary point process, Ripley's $K(r)$ function \citep{ripley2005spatial} is defined as the expected number of neighbors located within distance $r$ from a typical point, divided by the intensity of the process. Since some of the processes being modeled are non-stationary, we shall use $K_{inhom}$ \citep{baddeley2000non}, the inhomogeneous $K$-function. For either a homogeneous or inhomogeneous process $i$, we can estimate $K_i(r)$ by
\beq 
    \hat{K}_i(r) = \sum_{\bs, \bu \in Y_i} \frac{\mathbb{I}(0 < || \bs - \bu || \leq r)}{\lambda_i(\bs;\hat{\btheta}_i)\lambda_i(\bu; \hat{\btheta}_i)}w_{\bs,\bu}, \label{eq: Kfn}
\eeq 
where $\hat{\btheta}_i$ is the set of estimated parameters that control process $i$, and $w_{\bs,\bu}$ is the Ripley's edge-correction factor \citep{ripley1977modelling}. For homogeneous point patterns, the estimate (\ref{eq: Kfn}) is close to $\pi r^2$ for all values of $r$, and thus the estimated $K$-functions corresponding to $Y_j$, for $j=1,\ldots,p$, $p+q+1,\ldots,m$ should be close to $\pi r^2$. For inhomogeneous processes, the closed form expression is not evident, as is the case for processes $Y_j$ for $j=p+1,\ldots, p+q$.

By graphing the $K$-function, one can use it to assess homogeneity of processes. While it can thus be used to describe spatial patterns within a certain range, the $K$-function does not allow for quantitative validation of the point process models. Therefore, we interpret  the MCPP model based on the individual model parameters, which, in turn, influence the estimated $K$-functions. 

As outlined in Section \ref{s: model}, parameters of primary interest in the proposed model are the $\alpha_l$ and $h_l$ parameters, for $l = p+1, \ldots,  p+q$. The offspring density parameter $\alpha_l$ corresponds to how densely the $l$-th process clusters around its parent process, with higher values indicating denser clustering. The bandwidth parameter, $h_l$ determines how tightly the $l$-th process clusters around its parent process. A value of $h_l > h_0$ may indicate no or negligible clustering between the $l$-th process and its supposed parent process, where $h_0$ is a predetermined threshold based on domain knowledge from experts. Interpretation of $h_l$ can change depending on choice of $k_l( \cdot, \cdot)$. For example, in a Gaussian or Cauchy kernel, $h_l$ can be thought of as a scaled average distance between the offspring and parent, whereas in a uniform kernel, it represents the maximum distance between the parent and its offspring. Regardless of $k_l( \cdot, \cdot)$, one can also interpret $h_l$ as a parameter that estimates the median distance between the parents and offsprings: $\frac{32}{27}h$, $\sqrt{5}h$ and $\frac{2}{3}h$ for Gaussian, Cauchy and uniform kernels respectively. These parameters provide additional quantitative information about the relationship between a clustered process and its parent process, quantities not captured by the $K(r)$ function.
%\vspace{-2 em}
\subsection{Prior distributions and practical considerations} \label{s: prior}
We outline priors for the unknown model parameters to complete the Bayesian specification of the MCPP model. Specifically, we consider the following priors: 
\beq \label{eq: priors} 
\begin{split}
    \alpha_l &\iid \text{Gamma}(a_Y,b_Y) \,,\ h_l \iid \text{Half-Normal}(\sigma) \,,\ l=p+1, \ldots, p+q,\\
    \lambda^C_v &\iid \text{Gamma}(a_C,b_C) \,,\ v=1, \ldots, p\,,\ \lambda_j \iid \text{Gamma}(a,b) \,,\ j=p+q+1, \ldots, m,
\end{split} 
\eeq 
where $\iid$ denotes independent and identically distributed, and ($a,b,a_Y,b_Y,a_C,b_C,\sigma$) are hyperparameters to be specified. 

It is well appreciated that using a noninformative prior for spatial bandwidth or scale parameters is impossible due to numerical reasons, specifically, weak identifiability of the posterior distribution \citep{moller2003statistical, diggle2013statistical, kopecky2016bayesian}. We use a half-normal prior for the bandwidth parameters $h_l$; ensuring the prior assigns mass to relatively small positive values, stabilizing the parameter estimation. Specifically, the value of hyperparameter $\sigma$ can be set such that the 99-th percentile of the half-normal prior corresponds to the maximum distance $d$ between a parent point and offspring points. This choice helps sensitize the proposed Bayesian framework to stronger clustering within a small radius, which is expected when, for example, cells of two bacterial taxa directly interact. In general, choosing the value of $d$ should depend on the context.
%\vspace{-1 em}
\subsection{Computational scheme} 
\label{ss: comp}

Combining (\ref{eq: L}) and (\ref{eq: priors}), the joint posterior density for the proposed MCPP model can be written as $\pi(\btheta | Y) \propto L(Y|\btheta)\pi(\btheta)$, where $\pi(\btheta)$ is the product of all the individual prior densities. We then proceed to draw samples from the posterior distribution by using a Markov chain Monte Carlo (MCMC) algorithm. Conventional NSP-type approaches treat the number of parent points and their locations as random and thus require an additional reversible jump MCMC step to estimate the parameters associated with the latent parent process \citep{green1995reversible, moller2003statistical}. In contrast, the proposed framework exploits the fact that the parent processes are observed and proceeds without a complex birth-death-move algorithm. The components of $\btheta$ can then be updated by Gibbs sampling (exploiting conjugacies in the full conditionals) or via Metropolis-Hastings steps (details in Section C of the Supplementary Material).

One practical challenge is that the integral term in (\ref{eq: L}) does not have a closed-form expression. Therefore, we use Monte Carlo methods to approximate the integral for computational efficiency. In practice, the expression $\int_\mathcal{W} k_l(\bu - \bc_l, h_l) \text{d}\bu$ can be thought of as the probability of occurrence of $\mathbf{X}_l$ within the observation window $\mathcal{W}$, where $\mathbf{X}_l$ is a bivariate real-valued random variable with density $k_l(\cdot - \bc_l, h_l)$. Furthermore, in the proposed framework with Thomas processes, $k_l(\cdot - \bc_l,h_l)$ corresponds to a bivariate normal density function with mean $\bc_l$ and covariance matrix $h_l^2\mathbf{I}$. In this case, we draw samples from the bivariate normal distribution and compute the average proportion of points that fall within $\mathcal{W}$, which serves as an approximation to the integral term in (\ref{eq: L}). 

We further optimized the code by using the C language. An R package is available in the online repository at \url{https://github.com/SumanM47/MCPP.git}. The implementation of the model in the R package can generate 10,000 posterior samples in 1.5 minutes for a dataset with ${\sim} 150$ points from a single parent process with a total of ${\sim} 750$ points from two offspring processes on a Dell Latitude 7210 laptop with i5 
cores and 16 gigabytes of memory.

\subsection{A metric to explore model configurations} \label{s: DIC}

Using the MCPP model requires that we know or have hypotheses about the specific parent-offspring relationships present in the data. Complex multitaxon data can have many possible relationships present or absent, and users will not always have a specific \textit{a priori} hypothesis about the configurations. To illustrate this problem, we take up a simple example with 4 taxa labeled $A$, $B$, $C$ and $D$. There may be a parent-offspring relationship between $A$ (parent) and $B$ (offspring). We denote this by $A \rightarrow B$. The processes $C$ and $D$ may also be related similarly in addition to $A \rightarrow B$. We denote this combination as $A \rightarrow B \vdots C \rightarrow D$. If $C$ and $D$ were processes independent of each other and $A$ and $B$ were related, we would denote the model as $A \rightarrow B$ only. This means that when presenting all the relationships present in a model using this shorthand notation, we mention only the parent-offspring relationships to be modeled (separated by $\vdots$) and treat the other processes present in the data as independent of any other process. Additionally, there may be situations where two taxa, say $B$ and $C$ share the same parent $A$ as in i) in Section \ref{s: model}. We denote this configuration by $A \rightarrow BC$. If $B$ serves as an offspring to $A$ and as a parent to $C$ as in ii) in Section \ref{s: model}, we denote such a configuration as $A \rightarrow B \vdots B \rightarrow C$. These latter complicated relationships may or may not exist in the data and we can think of many more relations by permuting taxa's relative roles. In Section G of the Supplementary Material, we demonstrate how the configuration $A \rightarrow BC$ is formulated in the proposed model along with some others.

When applying to real data, it is impractical to assume that we know about all possible relationships present in the data. One may suspect their existence, or one may simply seek to explore consistency of the data with one or more possible clustering arrangements. For this purpose, we propose a model validation method to identify a best-fitting model among a series of models specifying different possible clustering relationships. Specifically, we use the DIC \citep{spiegelhalter2002bayesian} to compare models involving different parent-offspring structures and/or different kernels for the distribution of offspring around the parent. For example, models could be fit such that $A \rightarrow B$ is model 1, $C \rightarrow D$ is model 2, $A \rightarrow B \vdots C \rightarrow D$ is model 3, $A \rightarrow BC$ is model 4, $A \rightarrow B \vdots B \rightarrow C$ is model 5, and so on. The model with the smallest DIC would be identified as the best fit. This model comparison process enables use of the MCPP as an exploratory tool. Newly proposed clustering arrangements can be quantified and evaluated for consistency with the data.

%Use of the MCPP model requires prespecification of parent-offspring arrangement. In practice, one may lack prior knowledge of intertaxon spatial clustering. One may suspect their existence, or one may simply seek to explore consistency of the data with one or more possible clustering arrangements. For this purpose, we propose a model validation method to identify a best-fitting model among a series of models specifying different possible clustering relationships. Specifically, we use the DIC \citep{spiegelhalter2002bayesian} to compare models involving different parent-offspring structures and/or different kernels for the distribution of offspring around the parent. For example, models could be fit such that A is assumed to cluster around B in model 1, C around D in model 2, and both in model 3. The model with the smallest DIC would be identified as the best fit. This model comparison process enables use of the MCPP as an exploratory tool. Newly proposed clustering arrangements can be quantified and evaluated for consistency with the data.

\subsection{Goodness-of-fit} 

We also assess goodness-of-fit of the models by comparing the observed and estimated counts of different objects (in this case, taxa). Specifically, for the parent species, the estimated intensity parameter $\lambda^C_v$, $v=1, \ldots, p$, per unit area can be scaled by the area of the observation window. If the model is approximately correct, this quantity, $\lambda^C_v | W |$, should be close to the number of $v$-th parent cells counted in the observation window. For approximating the $l$-th offspring count, one can compute the posterior mean of $\alpha_l \sum_{\bc \in C_l} \int_W k_l(\bu - \bc, h_l) d\bu$ for $l=p+1, \ldots, p+q$. The integral term can be approximated by Monte Carlo methods as described in Section \ref{ss: comp}. This expression involves both the offspring density parameter and the bandwidth parameter and therefore helps validate the joint estimate of the $(\alpha_l,h_l)$ pair. \iffalse As the offspring density parameter summarizes the average number of offspring per parent cell, it follows that $\alpha\lambda| W |$ should produce a good approximation of the number of offspring that can be counted near parents cells. \fi In our example, $\lambda^C | W |$ should provide a good estimate of the number of \textit{Corynebacterium} cells; and  $\alpha_2 \sum_{\bc \in C_2} \int_W k_2(\bu - \bc, h_2) d\bu$, $\alpha_3 \sum_{\bc \in C_3} \int_W k_3(\bu - \bc, h_3) d\bu$ and $\alpha_4 \sum_{\bc \in C_4} \int_W k_4(\bu - \bc, h_4) d\bu$ should be close to the observed number of \textit{Streptococcus} clustered around \textit{Corynebacterium}, the observed number of \textit{Porphyromonas} clustered around \textit{Corynebacterium}, and the observed number of \textit{Pasteurellaceae} clustered around \textit{Streptococcus}, respectively.
%\vspace{-3 em}
\section{Simulation Studies}
\label{s: sim}
We performed two types of simulation studies to benchmark the performance of the method as an inference tool and as an exploratory tool. The first simulation study was designed to assess performance of the MCPP when the true parent-offspring relationships are known and to compare its performance to that of an existing method, the NSP. The second study was designed to assess the MCPP's performance in ranking the fit of models parameterizing different proposed parent-offspring relationships and different kernels, and in identifying the best-fitting model. In both types of studies, we compared the estimated and true parameter values under the generating model.
%\vspace{-1 em}
\subsection{Simulation I: Comparative performance for quantifying a prespecified configuration}

\subsubsection{Simulation set-up}

For simplicity, the unit square was taken as the observation window. We generated data under the model outlined in Section \ref{s: method}, with $q=2$ offspring taxa \emph{B} and \emph{C} around the same parent taxon \emph{A}, which was the only parent taxon ($p=1$). We considered twelve data scenarios (Supplementary Material Table S.1) that varied in terms of offspring density ($\alpha_2$, $\alpha_3$) being `Sparse', `Dense' or `Mixed'; bandwidth ($h_2$, $h_3$) being `Low' or `High'; and presence or absence of a taxon spatially unrelated to the multilayered arrangement (hereinafter referred to as ``unrelated" taxon). Throughout the scenarios, the intensity of the parent process ($\lambda_1^C$) and that of the process for the unrelated taxon ($\lambda_4$) were set to $150$ and $95$, respectively. For each of the twelve scenarios, we generated $100$ images, each analyzed as an independent dataset.
%\vspace{-1 em}
\subsubsection{Analysis plan and hyperparameters} \label{sss: sim1plan}

We applied the following approaches to each simulated dataset:
%\vspace{-0.5 em}
\begin{enumerate}[(i)]
    \item \emph{MCPP}: Multi-offspring taxa (\emph{A} as the parent of \emph{B} or \emph{C}) were jointly analyzed in one framework by using the MCPP.
    %\item \emph{MCPP-SO}: Two separate groups of taxa (\emph{A} as the parent of \emph{B}) and (\emph{A} as the parent of \emph{C}) were analyzed respectively using the MCPP.
    \item \emph{NSP}: Ignoring the parent taxon \emph{A}, only offspring taxa \emph{B} and \emph{C} were analyzed separately (by necessity, with the NSP). We applied the method of minimum contrast \citep{diggle2013statistical}, by using the R package \texttt{spatstat} \citep{baddeley2015R}.
\end{enumerate} 
%\vspace{-1.0 em}
The parameters estimated by the two approaches have different interpretations. For example, the bandwidth parameter $h_2$ in NSP models from analysis (ii) is interpreted as the distance scale for the unobserved cluster formed by offspring taxon \emph{B}, ignoring the observed parent taxon \emph{A}. Despite this distinction, we included analysis (ii) using the NSP because it is among the most relevant cluster point process models applied to this class of problems. In simulation studies, we primarily focused on the numerical performance of the methods in estimating the parameters rather than on their interpretations. We do not provide estimates of the $K$-functions because a) the data were generated under the homogeneity assumption and b) the $K$-function is a graphical inspection tool and cannot be used to summarize performance over multiple datasets.

For analysis (i), we set the hyperparameters ($a_Y$, $b_Y$, $a_C$, $b_C$, $a$, $b$) to $0.01$. We set the hyperparameter $\sigma$ to $0.02$ so that the $99$-th percentile of the prior distribution of $h_l$'s was approximately $0.05$ (i.e. $5\%$ of the length of the observation window). We used the posterior means as point estimates of the model parameters.

We estimated the offspring intensity ($\alpha_2, \alpha_3$) and bandwidth parameters ($h_2, h_3$) for the offspring processes (taxa \emph{B} and \emph{C}) and the intensity parameter ($\lambda_1^C$) for the parent process (taxon \emph{A}). For the MCPP, the presented statistics are the average of the posterior mean for the parameters for each of the datasets in a given scenario. We also computed the posterior standard deviation (SD) averaged over the $100$ datasets for each scenario and the standard deviations of the posterior means of the estimates for the different datasets (SD$_{EST}$). With the NSP, no uncertainty measure is available for the individual estimates for each dataset, so we computed only the standard deviation of each estimate across the datasets. For each scenario, we report the percentage of datasets in which the NSP failed to converge and computed average estimates and SEs based only on the datasets in which the model converged.
%\vspace{-1 em}
\subsubsection{Primary results}

The MCPP method performed well in estimating the true parameter values, with both a small SD and a small SD$_{EST}$ (Tables \ref{tab:sim_tab1} and S.2 of Supplementary Materials). The NSP method often failed to converge, in up to 70$\%$ of datasets, depending on the scenario. When the NSP did converge, it produced results that were nonsensical, with SEs too high to give any credibility to these estimates. 

 For both methods, standard errors increased in the high-bandwidth scenarios compared with the low-bandwidth scenarios. The parent process intensities were also captured better by the MCPP than by the NSP, especially in high-bandwidth scenarios. Lastly, performance of the MCPP approach was not affected by the presence of a taxon spatially unrelated to the multilayered arrangement (Tables \ref{tab:sim_tab1} and S.2 of Supplementary Materials). 
\subsubsection{Sensitivity analyses}

As explained in Section \ref{s: prior}, we used a half-normal prior for the bandwidth parameter. We conducted comprehensive sensitivity analyses (Supplementary Materials, Section E) to examine sensitivity of conlusions to the choice of prior distribution for the bandwidth parameter: half-normal, uniform, or log-normal. In summary, the proposed framework was not sensitive to the choice of the prior distribution for the bandwidth parameter in scenarios with low bandwidth. In high-bandwidth scenarios, however, elicitation of an informative prior seemed to help the proposed framework become more numerically stable, increasing precision. 

%\vspace{-0.5 em}
\subsection{Simulation II: Evaluating model selection}\label{sec:sim2}

\subsubsection{Simulation set-up}

The proposed MCPP model requires specification of the relationships one wishes to quantify and the offspring distribution kernel. This simulation study was designed to assess identification of the true model and to compare estimated parameters to the true model parameters. The unit square was taken as the observation window. We generated data under the model outlined in Section \ref{s: method}, with $q=2$ offspring taxa \emph{B} and \emph{C} around the same parent taxon \emph{A}, which was the only parent taxon ($p=1$), and $m=4$ by producing an extra unrelated taxon \emph{D}. We considered 4 different scenarios that differed in the choice of $k_l( \cdot, \cdot)$ (Gaussian or Cauchy) and in the magnitude of $h_l$ ($0.01$ or $0.05$). We generated $100$ datasets for each of the four scenarios. The intensity functions for \emph{A} and \emph{D} were set to $150$ and $100$, respectively, and ($\alpha_1$, $\alpha_2$) was set to $(2,1.5)$ in all the scenarios. Since the bandwidth parameters most strongly affect inference (Table \ref{tab:sim_tab1}), we focus our analysis on scenarios with varying $h_l$.
%\vspace{-1.5 em}
\subsubsection{Analysis plan and hyperparameters} \label{sss: anaII}

%We represent a parent-offspring relationship between taxa \emph{A} and \emph{B} by $A \rightarrow B$, with the arrow directed from the parent towards the offspring. If parent \emph{A} has two offsprings \emph{B} and \emph{C}, it is denoted by $A \rightarrow BC$. Different parent-offspring relationships within the same model are separated by $\vdots$. $A \rightarrow BC \vdots D \rightarrow E$ denotes that  \emph{A} is a parent to \emph{B} and \emph{C} and \emph{D} a parent to \emph{E}. The parent-offspring relationships in the true model are represented as $A \rightarrow BC$. Any taxa not represented in the notation and present in the model are modeled as independent HPP (e.g., taxon \emph{D}). We denote the kernels used for the model, whether Gaussian ($k^{(G)}$), Cauchy ($k^{(C)}$) or Uniform ($k^{(U)}$), by hyphenation. $A \rightarrow BC$-$k^{(G)}$ and $A \rightarrow BC$-$k^{(C)}$ denote the two true models for this simulation.

The parent-offspring relationships in the true model are represented as $A \rightarrow BC$ following the notation outlined in Section \ref{s: DIC}. Any taxa not represented in the notation and present in the model are modeled as independent HPP (e.g., taxon \emph{D}). We denote the kernels used for the model, whether Gaussian ($k^{(G)}$), Cauchy ($k^{(C)}$) or Uniform ($k^{(U)}$), by hyphenation. $A \rightarrow BC$-$k^{(G)}$ and $A \rightarrow BC$-$k^{(C)}$ denote the two true models for this simulation.

We generated data from the two true models under low and high bandwidth parameter settings. For each of these four types of scenarios, we fit 9 different models for each datatset by varying the parent-offspring relationships presented in the model ($A \rightarrow BC$, $A \rightarrow B$, and $A \rightarrow B \vdots D \rightarrow C$) and the choice of kernels ($k^{(G)}$, $k^{(C)}$, and $k^{(U)}$). For each scenario, the model that most often had the lowest DIC was considered the best fit. We obtained parameter estimates based on this model.

We set the hyperparameters $a_Y$, $b_Y$, $a_C$, $b_C$, $a$, $b$, and $\sigma$ to the same value as in Section \ref{sss: sim1plan} and estimated the same quantities of interest in the same manner.
%\vspace{-1 em}
\subsubsection{Primary results}

Use of the DIC successfully identified true parent-offspring arrangements for all datasets in all scenarios. Regarding kernel selection, in some scenarios the approach did not select the correct kernel used to generate the data (Table \ref{tab:sim_est}). This was particularly so with high bandwidth parameters. This discrepancy did not greatly affect the method's performance, which was excellent for all four data-generating mechanisms (Table \ref{tab:sim_est}). However, the estimated median distances between the parent offspring ($\frac{32}{27}h$ for Gaussian and $\sqrt{5}h$ for Cauchy) are similar under the scenarios with high bandwidth parameters, indicating the robustness of MCPP to the choice of kernel.

\section{Analysis of Human Microbiome Biofilm Image Data}
\label{s: data_app}

As another way to illustrate use of the proposed MCPP, we analyzed the microbial biofilm image data described in Section \ref{s: data_des}. One challenge of biofilm image data is higher-order spatial structure, such as is evident in the image of a dental plaque biofilm community. Among other reasons for macro-level community structure, the environment at the outer edge of dental biofilm differs from that near the tooth surface. More locally, \textit{Streptococcus} and \textit{Porphyromonas} cluster around \textit{Corynebacterium} only at its tip, not along its length and not at its base \citep[Figure \ref{fig:G1_CSP_SPo} and][]{welch2016biogeography}. Image sampling and processing further contribute to heterogeneity, especially in that the image is of a two-dimensional slice from a three-dimensional structure. Some areas of the image have a higher concentration of cross-sections in which \emph{Corynebacterium} cells themselves appear spherical, whereas other areas display large numbers of lengthwise \emph{Corynebacterium} filaments. Though some relationships are well known, there are undoubtedly intertaxon relationships that are yet unexplored and undocumented. Our method presents an opportunity to explore different possible relationships. With these objectives in mind, we performed two different analyses on this sample. In the first, we considered only the parent-offspring-type clustering configurations that have already been established. \citep{jones1972special,welch2016biogeography,morillo2022corncob} We analyzed data from the whole image and data from the same image divided into four quadrants, each with more homogeneous spatial patterns than the whole. In the second analysis, we explored various possible clustering configurations along with those considered in the first analysis; we evaluated consistency of these new structures with the data, i.e. their plausibility based on model fit. 

\subsection{Quantifying established relationships in dental plaque biofilm image data} \label{ss: whole_data}

\subsubsection{Analysis of whole image data}
%\subsubsubsection{Hyperparameters and analysis settings}
\paragraph{Hyperparameters and analysis settings}

From the $m=9$ different taxa probed in the human dental plaque sample, we analyzed the data for locations of three offspring taxa, \emph{Streptococcus} (S), \emph{Porphyromonas} (Po) and \textit{Pasteurellaceae} (Pa). The parent taxon for the first two taxa is \emph{Corynebacterium} (C), and for the third taxon, it is \emph{Streptococcus} which itself is an offspring taxon (Figure \ref{fig:multilayer_diag}). According to the notation introduced in Section \ref{s: model}, the process $Y_1$ corresponds to \emph{Corynebacterium}, which functions only as a parent taxon ($p=1$). $Y_2$, $Y_3$, and $Y_4$ represent the three offspring processes ($q=3$) for \emph{Streptococcus}, \emph{Porphyromonas}, and \emph{Pasteurellaceae}, respectively. The corresponding parent processes are $C_2= Y_1$, $C_3 = Y_1$ (\emph{Corynebacterium} as a parent), and  $C_4 = Y_2$ (\emph{Streptococcus} as a parent). The remainder of the taxa were modeled as HPPs. We denote the model as $C \ \rightarrow SPo {\vdots} S \rightarrow Pa$, by using the model notation introduced in Section \ref{sss: anaII} and the shorthand symbols for the taxa presented earlier here. We use the notations from Section \ref{s: model} to identify parameters to taxa and the notation from Section \ref{sss: anaII} to refer to the whole model. This convention has been used throughout the rest of the paper.

In the MCPP analyses, we set the hyperparameters at $a_Y=b_Y=a_C=b_C=a=b=0.01$ and $\sigma$ at $0.97$, such that the $99$th percentile for the bandwidth parameters was approximately $2.5 \,\ \mu$m. The inclusion of black space, where no taxa are observed, can deflate density estimates and induce spurious spatial correlations. We minimized unnecessary black space by using a convex hull of the observed locations as the analysis window (Figure \ref{fig:G1_comp}(b)). We also applied the NSP model individually on the three offspring processes, \emph{Streptococcus}, \emph{Porphyromonas} and \emph{Pasteurellaceae}.

\paragraph{Results}
For the MCPP, we ran three chains of length 3 million each with the initial 2 million iterations discarded as burn-in samples and the remaining samples thinned to obtain 10,000 posterior samples from each chain. The mixing was good for all the parameters. The estimates are posterior means and standard deviations based on the samples from three chains. The NSP model was run separately for the three offspring taxa by using the \texttt{thomas.estK} function in the R package \texttt{spatstat}. It uses the minimum contrast method to estimate the parameters and does not return any uncertainty estimates.

The MCPP identified the previously described multilayered arrangement in which \emph{Streptococcus} and \emph{Porphyromonas} cluster around \emph{Corynebacterium} and \emph{Pasteurellaceae} around \emph{Streptococcus}. The MCPP-based bandwidth estimates for \emph{Streptococcus} and \emph{Porphyromonas} clustering around \emph{Corynebacterium} were higher than expected, both above $\sim 10 \mu$m, and for \emph{Pasteurellaceae} around \emph{Streptococcus} it is $4.55 \mu$m (Table \ref{tab:res_tab_whole}). The NSP estimates for the bandwidth parameters involving \emph{Corynebacterium} are lower, while the corresponding offspring density parameters are very high. The estimated parameters from the MCPP are consistent in the sense that the expected count exactly matches the observed counts when rounded to the nearest integer. See the Supplementary Material for other parameter estimates (Table S.3).

The estimated (homogeneous) $K$-function for \emph{Corynebacterium} was higher than the expected value of $\pi r^2$ (Figure S.3, Supplementary materials). The clustering behavior of \emph{Streptococcus} and \emph{Porphyromonas} were not evident from the $K$-function. \emph{Pasteurellaceae} displayed strong clustering behavior based on its $K$-function (Figure S.3, Supplementary Material).
%\vspace{-1 em}
% \begin{table}
%     \centering
%     \caption{Results of MCPP and NSP analyses of  human dental plaque biofilm data: estimates (EST) and uncertainty measures for the offspring density $(\alpha_2, \alpha_3, \alpha_4)$, bandwidth $(h_2,h_3,h_4)$, and parent process $(\lambda^C)$ parameters. For the MCPP model, the estimates are the posterior means, and SD is computed as the posterior standard deviation for each of the parameters. For the NSP model, the estimates are the output of the minimum contrast method. SD is not computed for NSP, as the method does not provide one.}
%     \label{tab:res_tab_whole}
%     \begin{tabular}{ccccccc}
%     \hline
%                 &           &   & \multicolumn{2}{c}{MCPP} && NSP\\
%          \cline{4-5}
%          \cline {7-7}
%          Type   & Taxa      &   & EST & SD                  && EST\\
%          \hline
%          &\emph{Streptococcus}&$\alpha_2$ &1.34 &0.05 &&64.12\\
%          &&$h_2$ &10.96 &0.49 &&10.19\\Offspring&\emph{Porphyromonas}&$\alpha_3$ &2.20 &0.06 &&57.07\\
%          &&$h_3$ &14.14 &0.53 &&5.74\\&\emph{Pasteurellaceae}&$\alpha_4$ &0.42 &0.02 &&16.01\\
%          &&$h_4$ &4.55 &0.30 &&8.14\\
%          \hline         Parent&\emph{Corynebacterium}&$\lambda^C_1$ &0.02 &$< 0.01$ && $<0.01$\\
%          \hline
%     \end{tabular}
% \end{table}
%\vspace{-2 em}
\subsubsection{Analysis of subsetted image data} \label{s: subdata_app}

\paragraph{Analysis plan}
We subset the image in four equal-sized quadrants (Figure S.2, Supplementary Material) and analyzed data from each quadrant independently. While this ad hoc approach may be insufficient for some applications, testing and applying more optimal subsetting methods are beyond the scope of this work. The fact that taxa's abundance varied across the quadrants allowed for some comparison of performance (Supplementary Materials, Table S.4). The hyperparameters were set similarly as before, and convex hulls were created for each of the quadrants. We also performed NSP-based analysis of the three offspring taxa separately for each quadrant.

\paragraph{Results}
 The estimated bandwidth parameters $h_2, h_3$ and $h_4$ for the four quadrants ranged between $7.6$--$11.2 \mu$m, $7.2$--$15.2 \mu$m and $3.8$--$4.6 \mu$m, respectively. The estimated offspring density parameters $\alpha_2, \alpha_3$ and $\alpha_4$ were between $1.0$--$2.2$, $1.8$--$5.0$ and $0.4$--$0.6$, respectively. The estimated intensity parameter for \emph{Corynebacterium} was $0.01$--$0.02$ per unit area (Table \ref{tab:res_main}). The estimates of intensity parameters for the other taxa varied among the quadrants (Supplementary Table S.5).

The estimates for the parent intensity and offspring density parameters varied among quadrants and among parent-offspring pairs. The estimates for $\lambda^C_1$ in each quadrant were consistent with the observed counts. For example, in the second quadrant, there were $219$ observed \emph{Corynebacterium} cells, and the estimated intensity parameter was $0.02$ per unit area, giving an expected count of $\approx 219$. For each of the three offspring taxa, the estimated counts matched the observed counts when rounded to the nearest integer. For example, the estimated taxon counts for \emph{Streptococcus}, \emph{Porphyromonas} and \emph{Pasteurellaceae} in the third quadrant compute to  $163$, $269$ and $76$ respectively, which match the corresponding observed counts (Supplementary Table S.4). 

 The estimated bandwidth parameters $h_2, h_3$ and $h_4$ also varied by quadrant, exhibiting consistent patterns for different offspring-parent pairs. The NSP analysis of each of the offspring processes resulted in very low estimated bandwidth parameter values  (Table \ref{tab:res_main}).
\subsection{Application of the MCPP to explore proposed relationships in biofilm image data}

\subsubsection{Exploration strategy}

Here, the goal was to investigate the plausiblity of newly proposed parent-offspring-like clustering relationships by comparing the model fit when they are included with the fit of the original model developed based on prior knowledge, described in Section \ref{ss: whole_data}. The latter included specific configurations of \emph{Corynebacterium}, \emph{Streptococcus}, \emph{Porphyromonas} and \emph{Pasteurellaceae}. We investigated two relationships that have been reported and not confirmed. First, we fit models to evaluate a potential clustering relationship involving \emph{Streptococcus} around \emph{Fusobacterium} (F) \citep{lancy1983corncob} (See Supplementary Figure S.8). Second, we fit models to evaluate different possible clustering configurations of  \emph{Fusobacterium} with \emph{Leptotrichia} (L) (See Supplementary Figure S.9). Mark Welch et al. (2016)\cite{welch2016biogeography} also observed that the two genera tend to occupy the same area of the biofilm. How they may relate to each other spatially, if at all, is yet unexplored.
%\vspace{-1.5 em}
\subsubsection{Hyperparameters and analysis settings}\label{s: dic_models}

Using the notation in Section \ref{s: DIC}, we explored the following models:
%\vspace{-2 em}
\begin{multicols}{2}
\begin{enumerate}
    \item $C \rightarrow S Po \vdots S \rightarrow Pa$
    \item $C \rightarrow Po \vdots F \rightarrow S \vdots S \rightarrow Pa$
    \item $C \rightarrow S Po \vdots F \rightarrow S \vdots S \rightarrow Pa$
    \item $C \rightarrow S Po \vdots S \rightarrow Pa \vdots F \rightarrow L$
    \item $C \rightarrow S Po \vdots S \rightarrow Pa \vdots L \rightarrow F$
    \item $C \rightarrow S Po \vdots S \rightarrow Pa \vdots F \rightarrow L S$
    \item $C \rightarrow S Po \vdots S \rightarrow Pa \vdots L \rightarrow F \vdots F \rightarrow S.$
\end{enumerate}
\end{multicols}
%\vspace{-1 em}
In contrast to the models previously fit, the series of models in this substudy differ in the number of parent and offspring processes. The number of processes that serve only as parent ($p$) can be $1$ (Model 1) or $2$ (Models 2 through 7), while the number of unique offspring processes ($q$) can be $3$ (Models 1,2 and 3) or $4$ (Models 4 through 7). The choice for hyperparamaters $a_Y = b_Y = a_C = b_C = a = b = 0.01$ and $\sigma$ at $0.97$ remains the same with the same reasoning for their choice.
%\vspace{-1.5 em}
\subsubsection{Results}

\begin{figure}
    \centering
    \includegraphics[width=\linewidth]{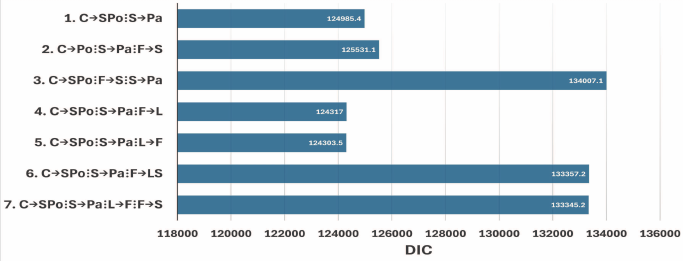}
    \caption{DIC values for analyzing the entire human dental plaque data using the 7 different models proposed in Section \ref{s: dic_models}. Lower values of DIC imply a better fit to the data. The abbreviations C, S, Po, Pa, F and L are used for \emph{Corynebacterium}, \emph{Streptococcus}, \emph{Porphyromonas}, \emph{Pasteurellaceae}, \emph{Fusobacterium} and \emph{Leptotrichia}. The $\rightarrow$ implies parent-offspring relationship with the arrow directed from the parent to the offspring(s) with seperate parent-offspring relationships separated by the $\vdots$ symbol (See Section \ref{s: DIC} for details of the notation used).}
    \label{fig:dic_comp}
\end{figure}

In this substudy involving serial fitting of models with different configurations, the lowest DIC values, $\sim$ 124,000, were from Models 4 and 5, configurations that included \emph{Fusobacterium} and \emph{Leptotrichia} in addition to the well-established configurations around \emph{Corynebacterium} (Figure \ref{fig:dic_comp}). These values were lower than $\sim$ 125,000, the DIC of the ``base model" that included only the previously known configurations (Model 1). The substudy provided little, if any, support for the proposed clustering of \emph{Streptococcus} around \emph{Fusobacterium} instead of \emph{Streptococcus} around \emph{Corynebacterium} (Model 2), as the DIC, $\sim$ 125,500, was higher than that for the base model. And, when \emph{Streptococcus} was assumed to cluster around both \emph{Corynebacterium} and \emph{Fusobacterium} (Models 3, 6, and 7), the DIC values, all $>$133,000, were by far the highest among the seven models (See Supplementary Table S.6 for the exact values).

The best-fitting model, Model 5 differed from the ``base model" only in the addition of a clustering relationship of \emph{Fusobacterium} and \emph{Leptotrichia}.  The estimates and the standard deviations for the relationships common to both models were nearly the same. The estimated number of \emph{Fusobacterium} cells per \emph{Leptotrichia} cell is $<$1, and the estimated bandwidth is roughly $5.22\,\ \mu$m (Table \ref{tab:explore_est}).
\section{Discussion and Conclusion}
\label{s: concl}

We have developed a multivariate model to make inference, simultaneously, about multiple parent-offspring clustering relationships, including multiple layers of clustering. The proposed MCPP framework produces model parameters that directly quantify multiple structural arrangements, in which locations of one type of object depend on locations of another, central object. This task cannot be achieved with a traditional NSP approach because it ignores the locations of central ``parent" objects. Because the MCPP assumes conditional independence, its use for modeling taxa individually will produce the same parameter estimates as the joint model. An advantage of the joint model over individual models is the ability to compare DIC values for parameterizations in which a set of taxa are modeled in different configurations. Indeed, we demonstrated that comparison of DIC values after serially fitting various MCPP model specifications is especially useful to explore possible clustering relationships when little is known \emph{a priori}.

In simulation studies, MCPP models correctly identified parent-offspring-type relationships and produced less bias and much lower empirical standard deviations for parameter estimates compared with NSP models, which produced rather nonsensical results. One seeming departure from ideal performance was that the MCPP did not always identify the kernel used to generate data, especially in high-bandwidth scenarios. This is not surprising because tail behaviors are similar for a low-bandwidth Cauchy kernel and a high-bandwidth Gaussian kernel. More importantly, MCPP performance was robust to the choice of kernel. 

Although development of the proposed MCPP model was motivated by oral microbiome biofilm image data, the approach is not specific to biofilms. The MCPP can be used in any image applications in which spatial dependencies reflect and provide information about the function of spatially arranged cells (or other objects) or about the function of arrangements themselves (e.g., the relationship between resprouters and seeders in a biodiverse plant community, as taken up by \cite{illian2009hierarchical}). %Such applications could be biomedical, nonclinical, or nonbiological.

We demonstrated feasibility and utility of the proposed method in application to biofilm image data that exhibit complex arrangements with nine taxa. Broadly speaking, the MCPP successfully captured the multilayered corncob-like structure among a group of four taxa. Additionally, application of the MCPP for data exploration provided evidence to support clustering between \emph{Fusobacterium} and \emph{Leptotrichia}. Because the DICs were nearly the same for models including either $L \rightarrow F$ or $F \rightarrow L$, the data support that they occupy the same physical niche in the biofilm rather than that one clusters around the other. We hope this observation will prompt further study of this previously unexplored relationship.

For some of the well-established clustering arrangements, the estimated bandwidth parameters were much greater than the approximately sub 5-micron distances expected from cell-to-cell (or nearly cell-to-cell) contact apparent in the visible corncob arrangements and likely when cells physico-chemically interact. Because of this seeming discrepancy, it could be tempting to conclude that the low parent-offspring bandwidths estimated by the NSP make the NSP a more valid and preferred approach than the MCPP. However, because the classical NSP model ignores the location of parent cells, it highly underestimated the number of cluster centers (``parent" cells). To compensate, it greatly overestimated the offspring density. Additionally, in ignoring parent locations, the NSP models self-clustering instead of parent-offspring clustering. Therefore, though seemingly appealing, the NSP produces estimates that are not appropriate to quantify the multilayered intertaxon relationships we sought to investigate.

It would also be inappropriate to interpret the MCPP’s high average estimated parent-offspring bandwidth strictly as bias or as reflecting a universal drawback to the proposed approach. Rather, the seeming discrepancy suggests the need for careful interpretation of model parameters from such a complex, heterogeneous image. It also highlights a need for modifications to the approach if a simple, direct interpretation about cell-to-cell (parent-to-offspring) contact is the goal. 

Two characteristics of the dental plaque biofilm data present particular challenges. First, cells of \emph{Corynebacterium} are filamentous. When the locations of imputed centroids are used to estimate average distances to neighboring cells of a different taxon, centroid-based methods may give misleading estimates of true cell-to-cell distances. This form of bias may be exacerbated by the specific biological organization, in that the spherical offspring taxa (\emph{Streptococcus} and \emph{Porphyromonas}) cluster around only one end of the parent \emph{Corynebacterium} filaments. Estimates of average \emph{Pasteurellaceae}-\emph{Streptococcus} distances were much smaller and closer to the expected range for cell-to-cell contact, likely because these clusters involved two types of similarly sized, spherical organisms.

One way to improve the MCPP performance, therefore, might be to model the shapes of cells by bi-axial spheroids \citep{clem1992towards} or elliptical cluster processes \citep{meinhardt2012modeling}. Another might be to use outline-based approaches, such as the Hausdorff distance \citep{huttenlocher1993comparing}, rather than relying on the geographic coordinates of imputed centroids to locate each filamentous cell. Nevertheless, such complicated adaptations of the MCPP approach are likely to be unnecessary if qualitative inference about the clustering arrangement suffices. Another option we are exploring is to condition models on features empirically identified in the image. Limiting the MCPP analysis only to parent or offspring cells within some short distance of each other (e.g., radius = 10 $\mu$m) might improve its usefulness in similar applications, compared with analyses yielding only marginal bandwidth estimates over a heterogeneous spatial structure. 

To explore another challenging feature, higher-order spatial structure, we analyzed subsetted image data. Between-quadrant variability in parameter estimates from both the MCPP and NSP methods suggests the subsetting approach was helpful. Even more powerful would be incorporation of the MCPP into a broader modeling framework that could capture higher-order spatial structure, for example, through regression parameters. The proposed model in (\ref{eq: all_taxa}) is flexible in that it can accommodate different standard modeling frameworks. For example, one can choose the form of $\lambda_i(\bs), i=1, \ldots, m$, and replace the HPP components in (\ref{eq: all_taxa}) by multivariate log-Gaussian Cox process components. Such an extension would enable the characterization and quantification of more complex spatial correlation structures among multiple taxa or other types of objects. We are currently developing these models.

We are also pursuing an avenue that will further enhance usefulness of the approach, by developing a meta-analytic framework to combine data from multiple images. In 100 sampled images of tongue biofilm from five donors, both within-sample and across-sample variability of inter-taxon spatial relationships is apparent \citep{wilbert2020spatial}. This variability can be quantified via a meta-analytic, multivariate, log-Gaussian Cox process model. To date, it has not been standard to apply formal unified models to combine data across multiple biomedical images. Instead, most practitioners have relied on post hoc comparisons, such as through ANOVA or non-parametric two-group comparisons. An efficient meta-analytic approach would also allow estimation and testing of covariate relationships with specific spatial structures, thereby increasing the scope and applicability of the proposed method.  

Validation of MCPP models is not straightforward. Complex spatial structures generally do not permit out-of-sample prediction or split-sample cross-validation. In the context of traditional Bayesian point process models, the empirical spatial distributions can be compared with those based on posterior predictive samples \citep{leininger2017bayesian}. However, this observed-versus-expected approach is challenged by the complexity of the data. Residuals for each of the sub-processes in (\ref{eq: all_taxa}) are easily obtained. Yet, a good match of observed and expected counts for one process can be misleading about overall model fit if another process is poorly estimated. This is exactly what happened with the NSP analysis of real data, where predicted offspring counts were accurate, and predicted parent counts were grossly underestimated. In contrast, the MCPP models produced nearly perfect prediction of counts of different taxa in the observation window. To our knowledge, there is as yet no valid method to combine the multiple residuals to produce a summary statistic reflecting overall goodness-of-fit. Using the DIC as a means to compare fit is valid only when the same method is used across the models.

We have proposed a novel MCPP method for simultaneously quantifying multilayer, multivariate spatial relationships that we applied to explore, confirm, and quantify spatial arrangements of microbial cells in dental plaque biofilm image data. The proposed method exploits information about locations of objects at the center of a cluster of unlike objects, providing distinct advantages over the classic NSP model when information about locations of central ``parent" objects is available. The MCPP model clearly outperformed the existing cluster point process model in every scenario in numerical studies.

\section*{Supplementary Material and Software}
Supplementary file with more images, tables and details about the method is supplied. R-package MCPP containing code to apply the method is available on request.
\ack
\section*{Acknowledgements}
The authors were funded by NIH grants GM126257, DE026872, DE027486, ES000002, DE016937 and DE022586. We are grateful for this support.

%\subsection{Bibliography}
%\nocite{*}% Show all bib entries - both cited and uncited; comment this line to view only cited bib entries;
\bibliography{WileyNJD-AMA}%

\clearpage

\begin{table}
    \centering
    %\spacingset{1.175}
    \caption{The true value, estimates (EST), and uncertainty measures for the offspring density ($\alpha_2$, $\alpha_3$), bandwidth ($h_2$, $h_3$), and parent process ($\lambda^C$) parameters from the MCPP and NSP analyses in the first six simulated scenarios (those without any unrelated taxon). For the MCPP model, the estimates are the posterior means averaged over different datasets, the SD is computed by averaging the posterior standard deviation over different datasets, and the SD$_{EST}$ is computed as the standard deviation of the estimates over the datasets. For the NSP model, the estimates are the outputs of the minimum contrast method, and SE is calculated similarly by using these estimates. The SD for the NSP model is not computed, as the method does not provide an uncertainty measure. The last column ($\%$F) refers to the percentage of datasets in which the NSP model failed to converge for a given scenario.}
    \label{tab:sim_tab1}
    \scalebox{1}{\begin{tabular}{cc rrrr r rrr}
    \hline
    &&True &\multicolumn{3}{c}{MCPP}& &\multicolumn{3}{c}{NSP}\\
    \cline{4-6} \cline{8-10}
    Scenario& &value& EST&SD& SD$_{EST}$& &EST&SE&$\%$F  \\
    \hline
         &$\alpha_2$&1.50&1.54&0.10&0.11&&2.34&7.71&  \\
         &$\alpha_3$&1.00&1.03&0.08&0.09&&1.97&6.84&  \\
         1&$h_2$&0.01&0.01&$<0.01$&$<0.01$&&0.01&0.04&6  \\
         &$h_3$&0.02&0.02&$<0.01$&$<0.01$&&0.70&6.43&  \\
         &$\lambda^C_1$&150.00&164.07&13.16&13.28&&170.78&52.51&  \\
         \hline &$\alpha_2$&1.50&1.50&0.11&0.11&&291.16&383.69&  \\ &$\alpha_3$&1.00&1.02&0.08&0.08&&1.06&0.40&  \\
         2&$h_2$&0.10&0.08&0.01&0.01&&8.33&23.82&26  \\
         &$h_3$&0.01&0.01&$<0.01$&$<0.01$&&0.01&$<0.01$&  \\ &$\lambda^C_1$&150.00&162.74&13.05&13.52&&679.72&2378.68&  \\
         \hline &$\alpha_2$&4.00&4.05&0.14&0.16&&13.30&66.75&  \\ &$\alpha_3$&3.00&3.06&0.13&0.11&&10.60&75.03&  \\
         3&$h_2$&0.01&0.01&$<0.01$&$<0.01$&&0.02&0.10&1  \\
         &$h_3$&0.02&0.02&$<0.01$&$<0.01$&&0.03&0.08&  \\ &$\lambda^C_1$&150.00&204.84&14.50&14.29&&209.22&49.75&  \\
         \hline &$\alpha_2$&4.00&4.01&0.17&0.17&&710.62&648.03&  \\ &$\alpha_3$&3.00&3.04&0.13&0.12&&3.00&0.62&  \\
         4&$h_2$&0.10&0.09&0.01&0.01&&1.21&0.77&56 \\
         &$h_3$&0.01&0.01&$<0.01$&$<0.01$&&0.01&$<0.01$&  \\ &$\lambda^C_1$&150.00&202.08&14.34&15.32&&20.33&45.79&  \\
         \hline &$\alpha_2$&4.00&4.05&0.15&0.16&&4.07&0.68&  \\ &$\alpha_3$&1.00&1.03&0.17&0.09&&3.06&14.91&  \\
         5&$h_2$&0.01&0.01&$<0.01$&$<0.01$&&0.01&$<0.01$&1  \\
         &$h_3$&0.02&0.02&$<0.01$&$<0.01$&&1.10&10.15&  \\ &$\lambda^C_1$&150.00&203.82&14.51&15.40&&203.47&35.16&  \\
         \hline &$\alpha_2$&4.00&4.02&0.17&0.14&&685.54&694.99&  \\ &$\alpha_3$&1.00&1.01&0.07&0.06&&2.49&10.25&  \\ 6&$h_2$&0.10&0.09&0.01&0.01&&1.48&1.21&53  \\ &$h_3$&0.01&0.01&$<0.01$&$<0.01$&&0.02&0.07&  \\ &$\lambda^C_1$&150.00&198.19&14.22&13.86&&35.99&99.33&  \\
         \hline \end{tabular}}
\end{table}

\begin{table}
    \centering
    \caption{Estimates (EST) and uncertainty in the estimates (SD and SD$_{EST}$) for the model parameters for the selected (best fit) model across four simulation scenarios outlined in Section \ref{sec:sim2}. Fitted models are represented by hyphenated combination of different arrangements ($A \rightarrow BC$, $A \rightarrow B$ or $A \rightarrow B {\vdots} D \rightarrow C$) and different offspring kernels ($k^{(G)}$, $k^{(C)}$ or $k^{(U)}$). Data generation scenario is denoted by hyphenated combination of offspring kernel ($k^{(G)}$ or $k^{(C)}$) and the arrangement ($A \rightarrow BC$) with the bandwidth magnitude (high: $h_l=0.05$ or low: $h_l=0.01$) specified separately. The last column (\%sel) signifies \% of datasets for which the best fit model had the lowest DIC and were used to get these results.}
    \begin{tabular}{cccrrrrr}
    \hline
        True model    & Selected    &   & True      &       &       &               &  \\
         (Bandwidth)  & model       &   & value     & EST   & SD    & SD$_{EST}$    & \%sel \\
         \hline
          & & $\alpha_2$ & 2 & 2.03 & 0.13 & 0.12 & \\
          & & $\alpha_3$ & 1.5 & 1.51 & 0.11 & 0.10 & \\
         $A \rightarrow BC$-$k^{(G)}$&$A \rightarrow BC$-$k^{(G)}$& $h_2$& 0.01& 0.01 & $<$0.01 & $<$0.01& 100\\
         (low)& & $h_3$ & 0.01 & 0.01 & $<$0.01 & $<$0.01 & \\
         & & $\lambda^C_1$ & 150 & 128.06 & 11.25 & 12.37 & \\
         \hline
         & & $\alpha_2$ & 2 & 2.01 & 0.13 & 0.13 & \\
          & & $\alpha_3$ & 1.5 & 1.52 & 0.11 & 0.13 & \\
         $A \rightarrow BC$-$k^{(G)}$&$A \rightarrow BC$-$k^{(G)}$& $h_2$& 0.05& 0.05 & $<$0.01 & $<$0.01& 81$^\dag$\\
         (high)& & $h_3$ & 0.05 & 0.05 & $<$0.01 & 0.01 & \\
         & & $\lambda^C_1$ & 150 & 126.36 & 11.17 & 11.70 & \\
         \hline
         & & $\alpha_2$ & 2 & 2.00 & 0.13 & 0.13 & \\
          & & $\alpha_3$ & 1.5 & 1.50 & 0.11 & 0.10 & \\
         $A \rightarrow BC$-$k^{(C)}$&$A \rightarrow BC$-$k^{(C)}$& $h_2$& 0.01& 0.01 & $<$0.01 & $<$0.01& 100\\
         (low)& & $h_3$ & 0.01 & 0.01 & $<$0.01 & $<$0.01 & \\
         & & $\lambda^C_1$ & 150 & 127.23 & 11.21 & 11.04 & \\
         \hline
         & & $\alpha_2$ & 2 & 2.00 & 0.15 & 0.14 & \\
          & & $\alpha_3$ & 1.5 & 1.50 & 0.13 & 0.12 & \\
         $A \rightarrow BC$-$k^{(C)}$&$A \rightarrow BC$-$k^{(C)}$& $h_2$& 0.05& 0.05 & 0.01 & 0.01& 94$^\ast$\\
        (high)& & $h_3$ & 0.05 & 0.05 & 0.01 & 0.01 & \\
         & & $\lambda^C_1$ & 150 & 128.87 & 11.29 & 10.44 & \\
         \hline
    \end{tabular}
    \label{tab:sim_est}
    
    \raggedright{\footnotesize{$^\dag$ The configuration $A \rightarrow BC$ was correctly selected for all datasets while the choice of $k^{(C)}$ yielded the best fit for 19\% of datasets. \\
    $^\ast$ The configuration $A \rightarrow BC$ was correctly selected for all datasets while the choice of $k^{(G)}$ yielded the best fit for 6\% of datasets.\\
    NOTE: Throughout values are based on the model selected most time among the 100 datasets.}}
\end{table}

\begin{table}
    \centering
    \caption{Results of MCPP and NSP analyses of  human dental plaque biofilm data: estimates (EST) and uncertainty measures for the offspring density $(\alpha_2, \alpha_3, \alpha_4)$, bandwidth $(h_2,h_3,h_4)$, and parent process $(\lambda^C)$ parameters. For the MCPP model, the estimates are the posterior means, and SD is computed as the posterior standard deviation for each of the parameters. For the NSP model, the estimates are the output of the minimum contrast method. SD is not computed for NSP, as the method does not provide one.}
    \label{tab:res_tab_whole}
    \begin{tabular}{ccccccc}
    \hline
                &           &   & \multicolumn{2}{c}{MCPP} && NSP\\
         \cline{4-5}
         \cline {7-7}
         Type   & Taxa      &   & EST & SD                  && EST\\
         \hline
         &\emph{Streptococcus}&$\alpha_2$ &1.34 &0.05 &&64.12\\
         &&$h_2$ &10.96 &0.49 &&10.19\\Offspring&\emph{Porphyromonas}&$\alpha_3$ &2.20 &0.06 &&57.07\\
         &&$h_3$ &14.14 &0.53 &&5.74\\&\emph{Pasteurellaceae}&$\alpha_4$ &0.42 &0.02 &&16.01\\
         &&$h_4$ &4.55 &0.30 &&8.14\\
         \hline         Parent&\emph{Corynebacterium}&$\lambda^C_1$ &0.02 &$< 0.01$ && $<0.01$\\
         \hline
    \end{tabular}
\end{table}

\begin{table}
    \centering
    %\spacingset{1.175}
    \caption{Results of MCPP and NSP analyses of  human dental plaque biofilm data: estimates (EST) and uncertainty measures for the offspring density $(\alpha_2, \alpha_3, \alpha_4)$, bandwidth $(h_2,h_3,h_4)$, and parent process $(\lambda^C)$ parameters from separate analyses of data from the four quadrants of the image. For the MCPP model, the estimates are the posterior means, and SD is computed as the posterior standard deviation for each of the parameters. For the NSP model, the estimates are the output of the minimum contrast method. SD is not computed for NSP, as the method does not provide one.}
    \label{tab:res_main}
    \begin{tabular}{ccc c rr r r}
    \hline
    &&&&\multicolumn{2}{c}{MCPP}&&NSP\\
    \cline{5-6}
    \cline{8-8}
    Segment&Type&Taxa&&EST&SD&&EST\\
    \hline
    &&\emph{Streptococcus}&$\alpha_2$&2.22&0.22&&9.94\\
    &&&$h_2$&8.32&0.56&&2.88\\
    &Offspring&\emph{Porphyromonas}&$\alpha_3$&5.05&0.34&&24.19\\
    I&&&$h_3$&7.23&0.50&&2.55\\
    &&\emph{Pasteurellaceae}&$\alpha_4$&0.57&0.08&&15.39\\
    &&&$h_4$&3.82&0.40&&9.39\\
    \cline{2-8}
    &Parent&\emph{Corynebacterium}&$\lambda^C_1$&0.01&$<0.01$&&$<0.01$\\
    \hline
    &&\emph{Streptococcus}&$\alpha_2$&1.98&0.10&&41.89\\
    &&&$h_2$&11.23&0.58&&7.07\\
    &Offspring&\emph{Porphyromonas}&$\alpha_3$&2.91&0.13&&35.43\\
    II&&&$h_3$&15.21&0.79&&4.57\\
    &&\emph{Pasteurellaceae}&$\alpha_4$&0.36&0.03&&10.12\\
    &&&$h_4$&4.23&0.46&&6.72\\
    \cline{2-8}
    &Parent&\emph{Corynebacterium}&$\lambda^C_1$&0.02&$<0.01$&&$<0.01$\\
    \hline
    &&\emph{Streptococcus}&$\alpha_2$&1.04&0.08&&9.34\\
    &&&$h_2$&7.62&0.71&&3.36\\
    &Offspring&\emph{Porphyromonas}&$\alpha_3$&1.80&0.11&&18.66\\
    III&&&$h_3$&9.83&0.72&&1.92\\
    &&\emph{Pasteurellaceae}&$\alpha_4$&0.52&0.06&&3.78\\
    &&&$h_4$&4.50&0.43&&2.67\\
    \cline{2-8}
    &Parent&\emph{Corynebacterium}&$\lambda^C_1$&0.02&$<0.01$&&$<0.01$\\
    \hline
    &&\emph{Streptococcus}&$\alpha_2$&1.22&0.08&&32.29\\
    &&&$h_2$&9.41&0.81&&5.84\\
    &Offspring&\emph{Porphyromonas}&$\alpha_3$&2.11&0.11&&51.56\\
    IV&&&$h_3$&11.09&0.72&&4.14\\
    &&\emph{Pasteurellacea}&$\alpha_4$&0.46&0.05&&9.93\\
    &&&$h_4$&4.62&0.42&&4.42\\
    \cline{2-8}
    &Parent&\emph{Corynebacterium}&$\lambda^C_1$&0.02&$<0.01$&&$<0.01$\\
    \hline
    \end{tabular}
\end{table}

\begin{table}
    \centering
    \caption{Parameter estimates (EST) and uncertainty quantification (SD) for the model best fit according to lowest DIC (Model 5). Parameter estimates are obtained by considering the posterior mean and SD computed by posterior standard deviation. $\alpha$ or $h$ subscripted by 3, 4, 5 and 6 refer to the offspring density and bandwidth parameters for \emph{Streptococcus}, \emph{Porphyromonas}, \emph{Pasteurellaceae} and \emph{Fusobacterium} respectively. The parameters $\lambda_1$ and $\lambda_2$ represent the intensity parameter for the two parent species \emph{Corynebacterium} and \emph{Leptotrichia} respectively.}
    \label{tab:explore_est}
    \begin{tabular}{cccrr}
    \hline
    Type & Taxon && EST & SD\\
    \hline
\multirow{8}{*}{Offspring}  &\emph{Streptococcus}   &$\alpha_3$ & 1.33 & 0.04\\
                            &                       &$h_3$ & 10.79 & 0.47\\
                            &\emph{Porphyromonas}   &$\alpha_4$ & 2.19 & 0.06\\
                            &                       &$h_4$ & 14.16 & 0.60\\
                            &\emph{Pasteurellaceae} &$\alpha_5$ & 0.42 & 0.02\\
                            &                       &$h_5$ & 4.33 & 0.26\\
                            &\emph{Fusobacterium}   &$\alpha_6$ & 0.57 & 0.02\\
                            &                       &$h_6$ & 5.22 & 0.32\\
\hline
\multirow{2}{*}{Parent}     &\emph{Corynebacterium} &$\lambda^C_1$ & 0.02 & $<0.01$\\
                            &\emph{Leptotrichia}    &$\lambda^C_2$ & 0.03 & $<0.01$\\
    \hline
    \end{tabular}
\end{table}

\end{document}

\title{Multivariate cluster point process to quantify and explore multi-entity configurations: Application to biofilm image data - Supplementary Materials}

\author
{Suman Majumder$^{1,*}$ , Brent~A.~Coull$^{2}$, Jessica~L.~Mark~Welch$^{3}$,\and
Patrick~J.~La Riviere$^{4}$, Floyd~E.~Dewhirst$^{3}$,\and
Jacqueline~R.~Starr$^{5,\dagger}$,Kyu~Ha~Lee$^{2,\dagger}$ \\ \\
$^{1}$ University of Missouri, Columbia, MO, USA\\
$^{2}$ Harvard T.H. Chan School of Public Health, Boston, MA, USA\\
$^{3}$ Forsyth Institute, Cambridge, MA, USA\\
$^{4}$ University of Chicago, Chicago, IL, USA\\
$^{5}$ Brigham and Women's Hospital, Boston, MA, USA\\
$^{\dagger}$ Co-senior authors\\}
%\email{sm8qr@missouri.edu}}

%\date{ }

%\volume{ }
%\pubyear{ }
%\artmonth{ }

%  The \doi command is where the DOI for your paper would be placed should it
%  be published.  Again, if you make one up and stick it here, it means 
%  nothing!

%\doi{ }

%  This label and the label ``lastpage'' are used by the \pagerange
%  command above to give the page range for the article.  You may have 
%  to process the document twice to get this to match up with what you 
%  expect.  When using the referee option, this will not count the pages
%  with tables and figures.  

%\label{firstpage}

\maketitle

\clearpage
\section{Neyman-Scott process}

 A Neyman-Scott process is a point process used for modeling parent-offspring clustering. In the simplest setting, consider the parent process $C$ to be a homogeneous Poisson point process with intensity $\lambda^C$. For each observation location $c \in C$, the cluster of offspring $Y_c$ is an independent Poisson process with intensity $\alpha k(\cdot - c,h)$, where $k(\cdot - c,h)$ is a probability distribution function parameterized by $h$ that determines the spread and distribution of the offspring locations around the parent $c$, and $\alpha > 0$ is the expected number of offspring per cluster. The Neyman-Scott process $Y$ is the union of all these offspring cluster processes, namely, $Y = \bigcup_{c \in C} Y_c$. Further details can be found in \cite{illian2008statistical} and \cite{chiu2013stochastic}, for example.
 
\clearpage
\section{Images of the taxa from the human dental plaque biofilm data not visualized in the image included in the main text}

\begin{figure}[!h]
    \centering
    \includegraphics[width=0.35\textwidth]{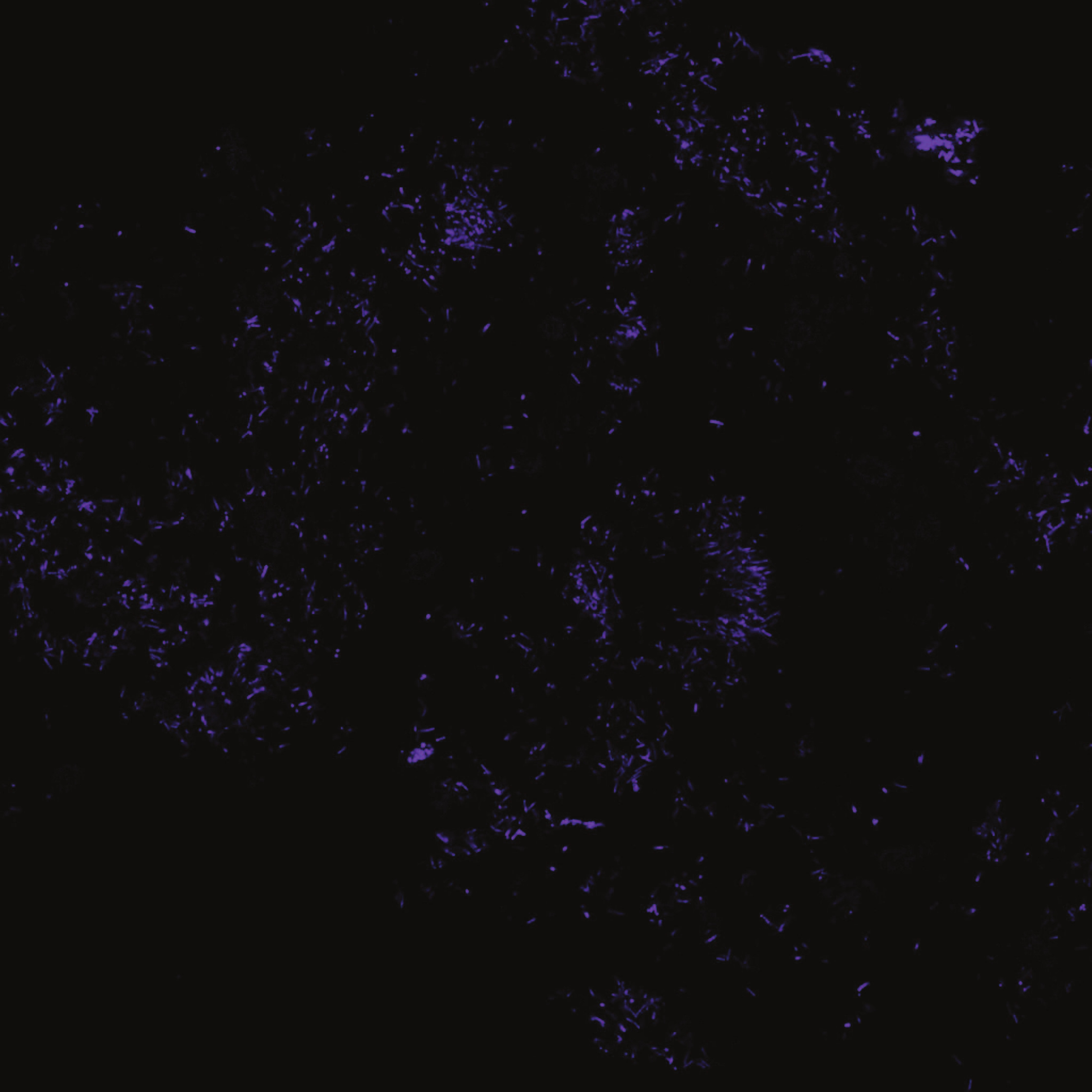} \includegraphics[width=0.35\textwidth]{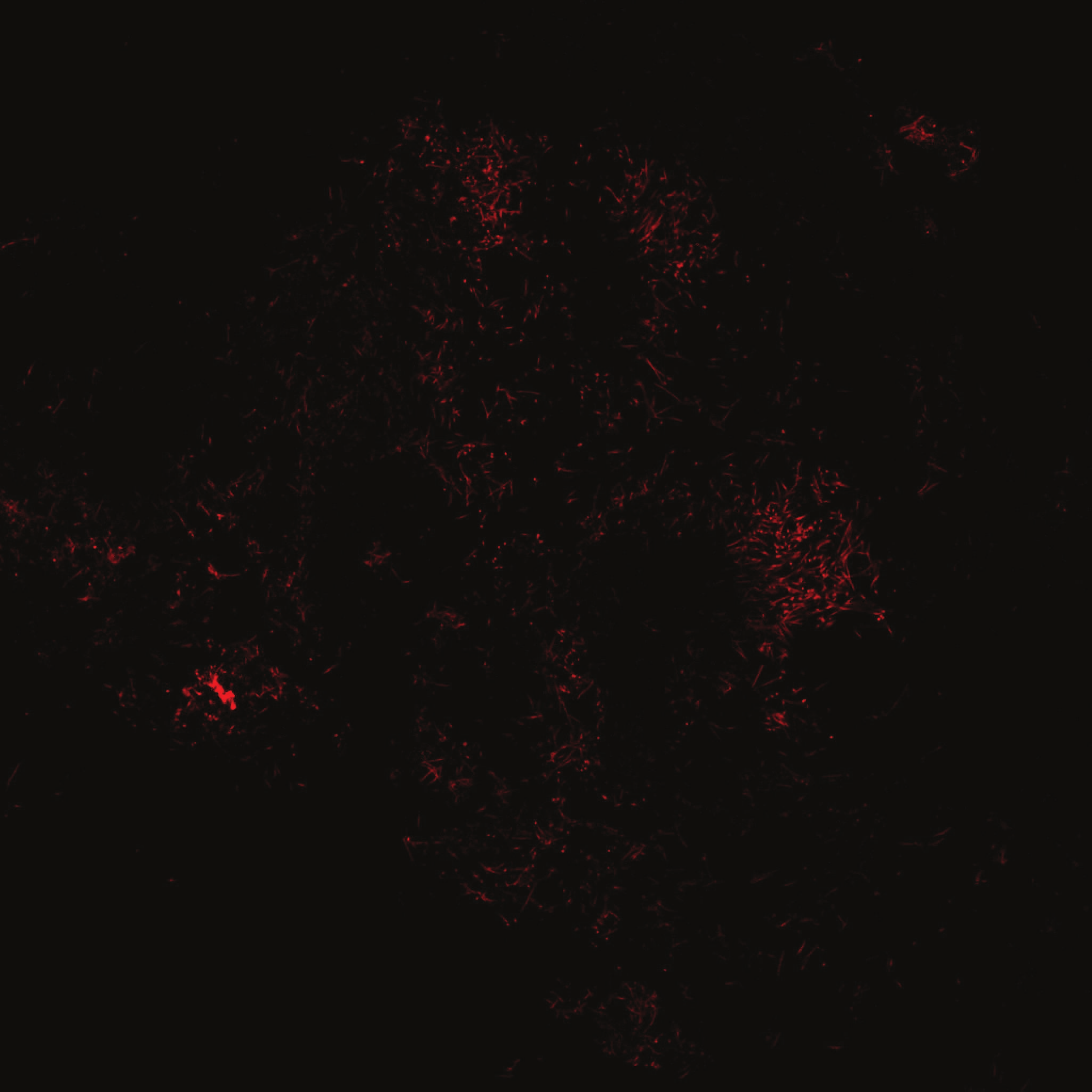}\\ \includegraphics[width=0.35\textwidth]{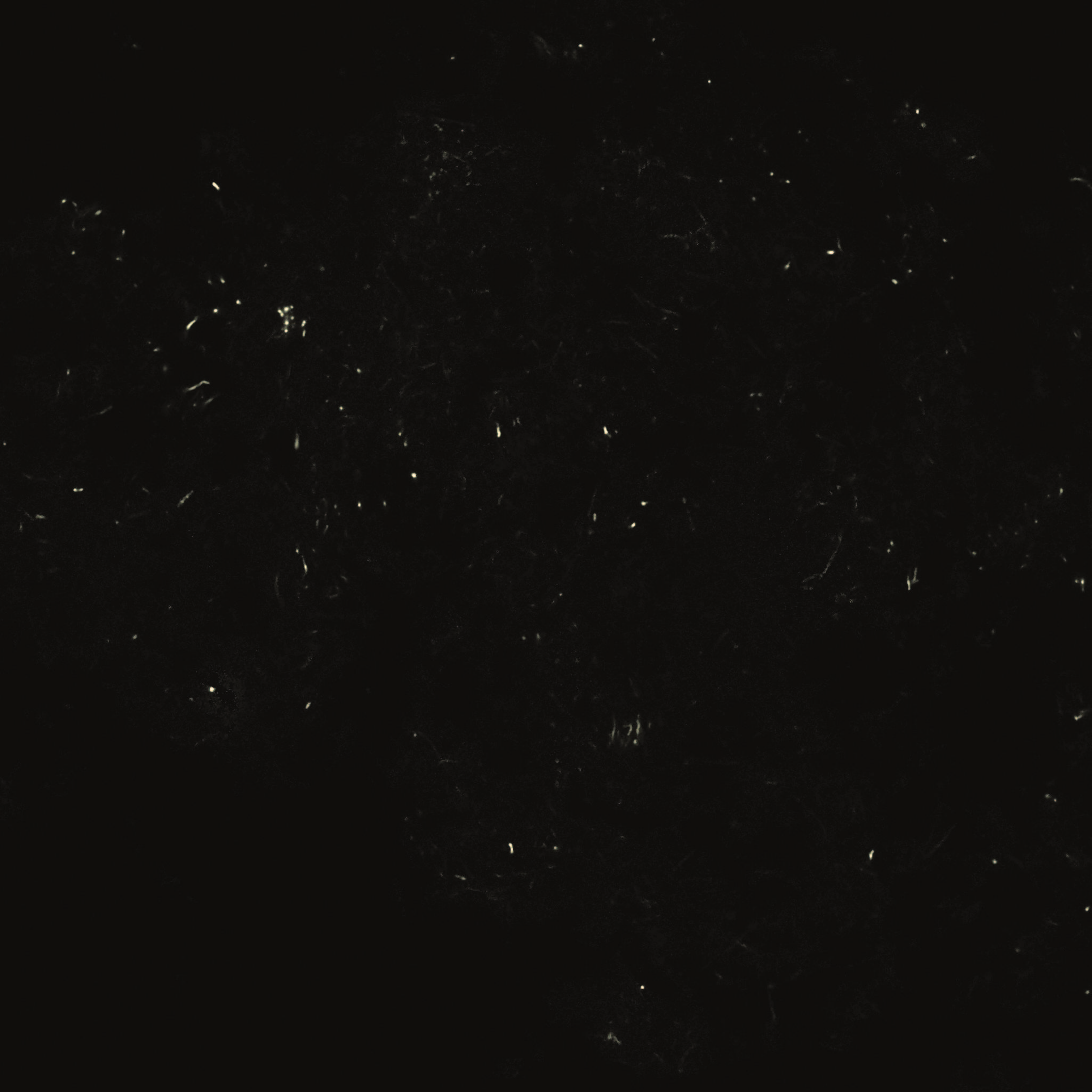}  \includegraphics[width=0.35\textwidth]{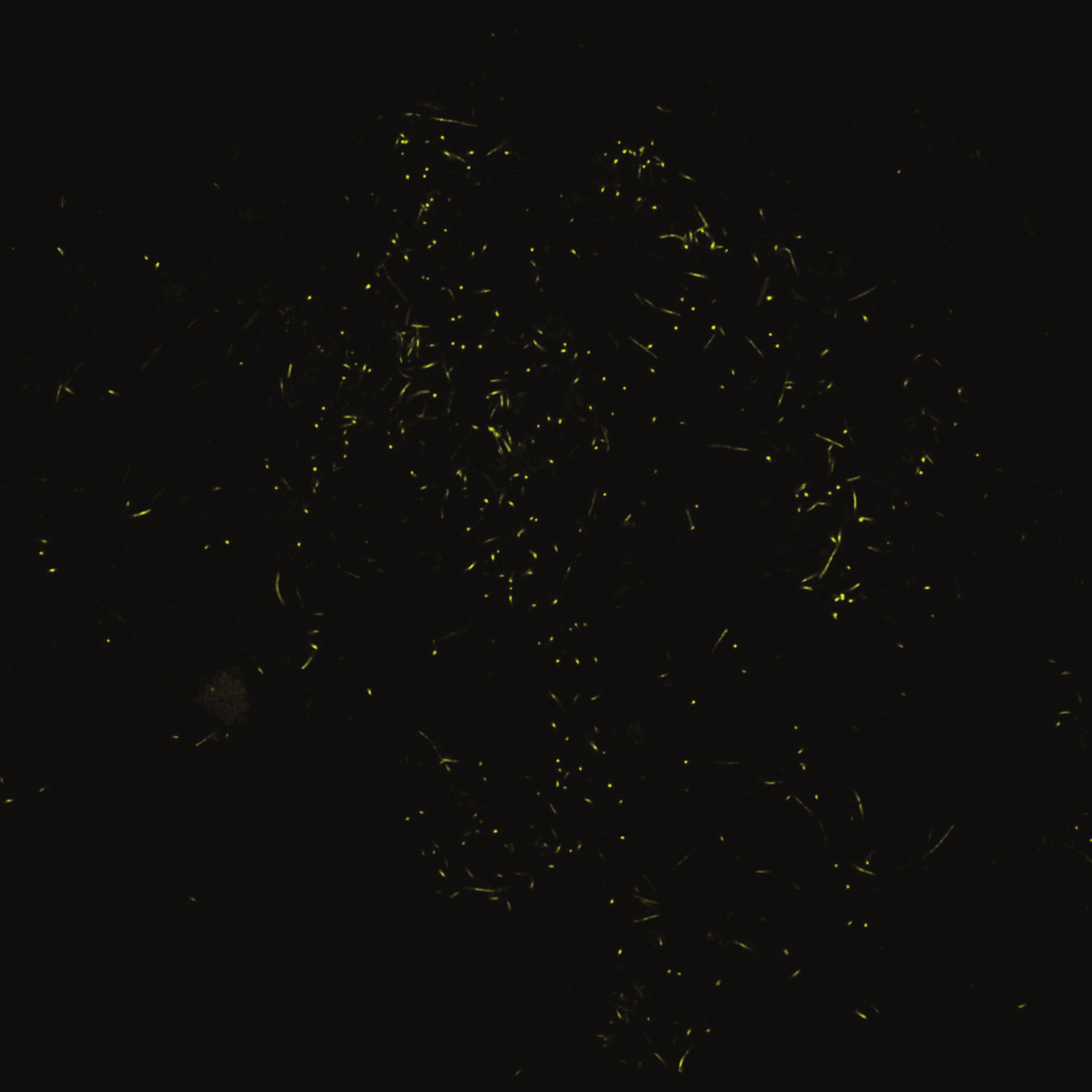}\\
    \includegraphics[width=0.35\textwidth]{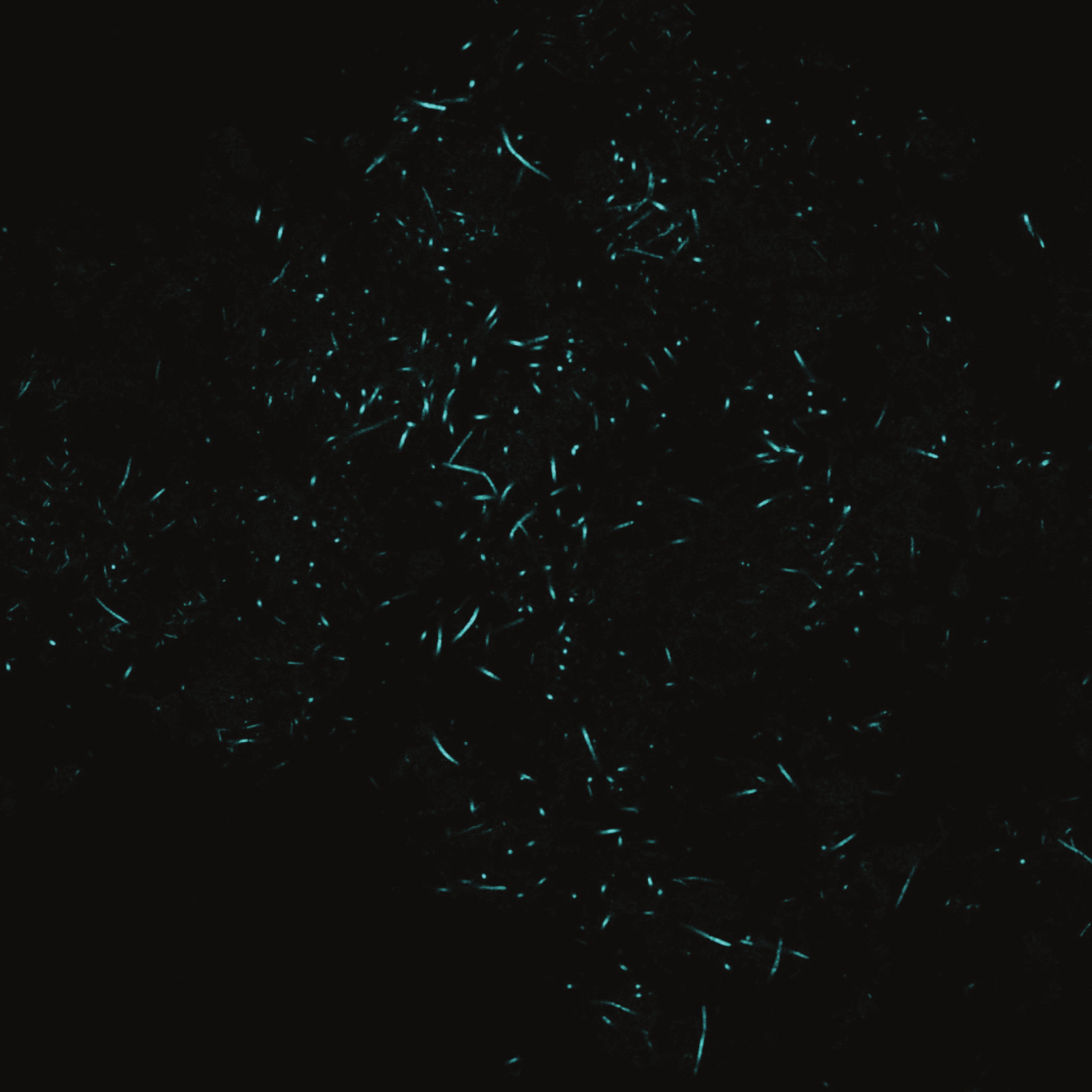} \includegraphics[width=0.35\textwidth]{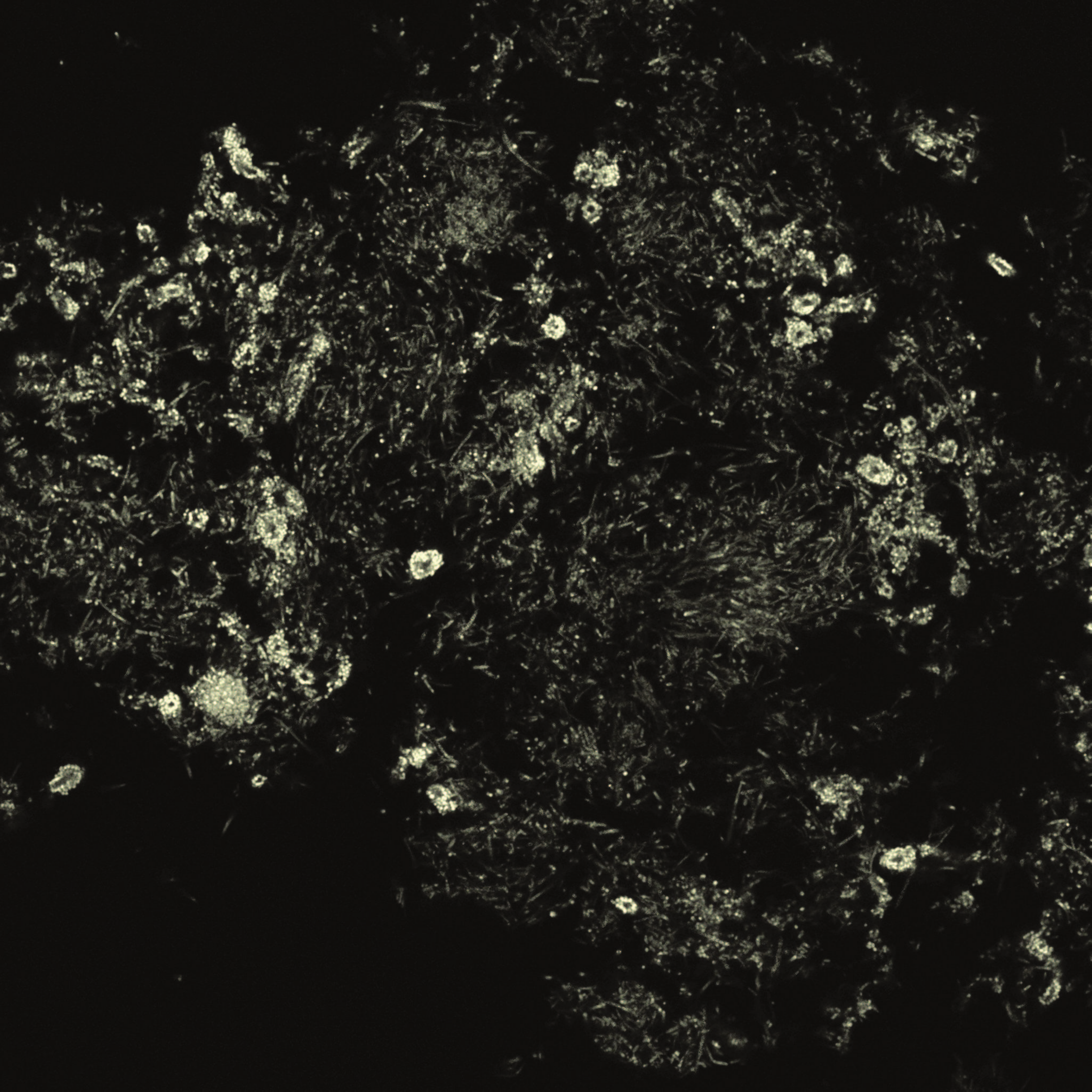}
    \caption{RGB images of \textit{Neisseriaceae} (top left), \textit{Capnocytophaga} (top right), \textit{Actinomyces} (middle left), \textit{Fusobacterium} (middle right), \textit{Leptotrichia} (bottom left) and \textit{Eubacterium} (bottom right) in the dental plaque biofilm sample. \textit{Eubacterium} denotes a probe for all oral bacteria. It is used for methodologic purposes to evaluate the completeness of the set of specific probes. Hence, it is omitted from analysis of community spatial structure. The genera shown here were modeled as homogeneous Poisson process in the data analysis. }
    \label{fig: other_taxa}
\end{figure}

\clearpage
\section{Computational details of the sampling algorithm}

We use a Markov chain Monte Carlo (MCMC) method to draw samples from the joint posterior distribution of $\btheta$. In the MCMC scheme, parameters are updated either by exploiting conjugacies inherent to the proposed model or by using a Metropolis-Hastings algorithm. 

\subsection{Updating parameters associated with offspring densities}

Let $\btheta^{-(\alpha)}$ denote a set of parameters $\btheta$ with $\alpha$ removed. The full conditional distribution for $\alpha_l \,,\ l=p+1,\ldots,p+q$ is 
\[
\alpha_l|\btheta^{-(\alpha_l)} \sim \mbox{ Gamma}(a_Y + n_l,b_Y + \sum_{\bc_l \in C_l} \int_W k_l(\bu - \bc_l,h_l) \,\ d\bu),
\] 
where $n_l$ is the number of observations of taxon $l$ in the window.

\subsection{Updating intensity parameters in homogeneous Poisson processes}

Posterior conjugacy is also achieved in the full conditional distributions of intensity parameters, $\lambda^C_v \,,\ v = 1, \ldots, p$ and $\lambda_j \,,\ j=p+q+1, \ldots, m$, which are given by
\[
    \lambda^C_v|\btheta^{-(\lambda^C_v)} \sim \mbox{ Gamma} (a_C + n_v, ~b_C + \lvert \mathcal{W} \rvert) \,,\ v=1, \ldots, p; \mbox{ and}
\]
\[
    \lambda_j|\btheta^{-(\lambda_j)} \sim \mbox{ Gamma} (a+n_j,~b+ \lvert \mathcal{W} \rvert) \,,\ j = p+q+1, \ldots, m;
\] where $n_v$ and $n_j$ are the numbers of observations for taxon $v$ and taxon $j$ within the window, respectively.

\subsection{Updating bandwidth parameters}

Since the full conditionals of the bandwidth parameters do not have standard forms, we use a random work Metropolis-Hastings step to update each of $h_l \,,\ l=1, \ldots, p$. Denote $h_j^{(t)}$ the sample for $h_j \,,\ j = p+1. \ldots, p+q$ from iteration $t$. For iteration $(t+1)$, we propose a candidate sample $h_j^*$ as a random draw from $N(h_j^{(t)},\sigma^2_{prop})$, where $\sigma^2_{prop}$ is the prespecified variance of the proposal density. The corresponding acceptance ratio computes to 
\[R = \frac{\exp \Lp -\alpha_l \sum_{\bc_l \in C_l} \int_W k(\bu - \bc_l,h_j^*) \,\ d\bu \Rp \prod_{\by \in Y_l} \Lp \sum_{\bc_l \in C_l} \int_W k(\bu - \bc_l,h_j^*) \Rp\exp \Lp -h_j^{*2}/2\sigma^2 \Rp \mathbb{I}(h_j^* > 0)}{\exp \Lp - \alpha_l \sum_{\bc_l \in C_l} \int_W k(\bu - \bc_l,h_j^{(t)}) \,\ d\bu \Rp \prod_{\by \in Y_l} \Lp \sum_{\bc_l \in C_l} \int_W k(\bu - \bc_l,h_j^{(t)}) \Rp\exp \Lp -h_j^{(t)2}/2\sigma^2 \Rp}.\] 
Then, we accept the proposed candidate $h_j^*$ as $h_j^{(t+1)}$ with probability $\mbox{min}\{R,1\}$ or keep $h_j^{(t+1)} = h_j^{(t)}$.

\pagebreak

\section{Additional Tables from the simulation study}

Here, we present additional details regarding the simulation scenarios (Table \ref{tab:sim_settings}) and results for the scenarios that included a taxon unrelated to the parent-offspring-type configurations of interest (Table \ref{tab:sim_tab2}). The presence of an unrelated taxon (Table \ref{tab:sim_tab2}) did not meaningfully affect the results (Table 1). Specifically, with or without this spatially unrelated taxon, the multivariate cluster point process (MCPP) performed better than the Neyman-Scott process (NSP) implementation in all aspects. The NSP often failed to converge, especially in scenarios where the bandwidth parameter was large.

\begin{table}[ht]
    \centering
    %\spacingset{1.175}
    \caption{A summary of twelve simulation scenarios considered in Section 4. The offspring density is controlled by setting $(\alpha_2,\alpha_3) = (1.5,1)$ for `Sparse', $(4,3)$ for `Dense' and $(4,1)$ for `Mixed' densities. Bandwidth `Low' sets $(h_2,h_3) = (0.01,0.02)$ and `High' to $(0.1,0.01)$. The setting ``Unrelated taxon" refers to whether there exists a taxon in the data spatially unrelated to the multilayered arrangement.}
    \label{tab:sim_settings}
    \begin{tabular}{cccc}
    \hline
        Scenario&Unrelated taxon&Offspring density&Bandwidth\\
        \hline
        1&Absent&Sparse&Low\\
        2&Absent&Sparse&High\\
        3&Absent&Dense&Low\\
        4&Absent&Dense&High\\
        5&Absent&Mixed&Low\\
        6&Absent&Mixed&High\\
        7&Present&Sparse&Low\\
        8&Present&Sparse&High\\
        9&Present&Dense&Low\\
        10&Present&Dense&High\\
        11&Present&Mixed&Low\\
        12&Present&Mixed&High\\
        \hline
    \end{tabular}
\end{table}

\begin{table}[]
    \centering
    \caption{The true value, estimates (EST), and uncertainty measures for the offspring density ($\alpha_2$, $\alpha_3$), bandwidth ($h_2$, $h_3$), and parent process ($\lambda^C_1$) parameters from the MCPP and NSP analyses in the last six simulated scenarios (those that included a spatially unrelated taxon). For the MCPP model, the estimates are the posterior means averaged over different datasets, the SD is computed by averaging the posterior standard deviation over different datasets, and the SD$_{EST}$ is computed as the standard deviation of the estimates over the datasets. For the NSP model, the estimates are the outputs of the minimum contrast method, and the SE is calculated similarly by using these estimates. The SD for the NSP model is not computed, as the method does not provide an uncertainty measure. The last column ($\%$F) refers to the percentage of datasets in which the NSP model failed to converge for a given scenario. There is no corresponding $\%$F column for the MCPP because all models converged.}
    \label{tab:sim_tab2}
    \begin{tabular}{c c rrrr r rrr}
    \hline
    &&True&\multicolumn{3}{c}{MCPP}&&\multicolumn{3}{c}{NSP}\\
    \cline{4-6}
    \cline{8-10}
    Scenario&&Value&EST&SD& SD$_{EST}$&&EST&SE&$\%$F  \\
    \hline
         &$\alpha_2$&1.50&1.53&0.10&0.10&&1.46&0.33&  \\ &$\alpha_3$&1.00&1.02&0.08&0.09&&3.31&20.25&  \\
         7&$h_2$&0.01&0.01&$<0.01$&$<0.01$&&0.01&$<0.01$&2  \\
         &$h_3$&0.02&0.02&$<0.01$&$<0.01$&&0.04&0.09&  \\ &$\lambda^C_1$&150.00&161.06&12.91&12.20&&171.35&34.72&  \\
         \hline &$\alpha_2$&1.50&1.48&0.11&0.09&&198.46&283.69&  \\ &$\alpha_3$&1.00&1.02&0.08&0.09&&0.98&0.28&  \\
         8&$h_2$&0.10&0.08&0.01&0.01&&10.30&28.72&36  \\
         &$h_3$&0.01&0.01&$<0.01$&$<0.01$&&0.01&$<0.01$&  \\ &$\lambda^C_1$&150.00&160.25&12.86&12.57&&939.70&2777.40&  \\
         \hline &$\alpha_2$&4.00&4.02&0.14&0.15&&8.77&48.78&  \\ &$\alpha_3$&3.00&3.05&0.13&0.13&&2.91&0.77&  \\
         9&$h_2$&0.01&0.01&$<0.01$&$<0.01$&&0.02&0.08&0  \\
         &$h_3$&0.02&0.02&$<0.01$&$<0.01$&&0.02&$<0.01$&  \\ &$\lambda^C_1$&150.00&202.78&14.38&14.29&&208.52&39.37&  \\
         \hline &$\alpha_2$&4.00&4.00&0.17&0.17&&613.34&569.20&  \\ &$\alpha_3$&3.00&3.02&0.13&0.14&&2.93&0.53&  \\
         10&$h_2$&0.10&0.09&0.01&0.01&&1.15&0.64&48  \\
         &$h_3$&0.01&0.01&$<0.01$&$<0.01$&&0.01&$<0.01$&  \\ &$\lambda^C_1$&150.00&200.49&14.26&14.48&&13.30&28.78&  \\
         \hline &$\alpha_2$&4.00&4.05&0.15&0.14&&18.45&87.48&  \\ &$\alpha_3$&1.00&1.00&0.07&0.08&&2.05&10.04&  \\ 11&$h_2$&0.01&0.01&$<0.01$&$<0.01$&&0.03&0.11&0  \\
         &$h_3$&0.02&0.02&$<0.01$&$<0.01$&&0.47&4.41&  \\ &$\lambda^C_1$&150.00&201.50&14.37&14.09&&203.36&53.30&  \\
         \hline &$\alpha_2$&4.00&4.02&0.17&0.17&&547.61&553.61&  \\ &$\alpha_3$&1.00&1.02&0.07&0.07&&0.97&0.24&  \\
         12&$h_2$&0.10&0.09&0.01&0.01&&1.04&0.82&70  \\
         &$h_3$&0.01&0.01&$<0.01$&$<0.01$&&0.01&$<0.01$&  \\ &$\lambda^C_1$&150.00&199.05&14.23&15.95&&26.62&41.46&  \\
         \hline
    \end{tabular}
\end{table}

\clearpage

\section{Sensitivity analyses regarding choice of prior for the bandwidth parameters }

As part of the simulation study described in Section 4, we also evaluated the MCPP method's sensitivity to choice of prior distribution for the bandwidth parameters. We considered four different prior distributions; 1) half-normal, 2) uniform, 3) log-normal with a flat tail and high variance and 4) log-normal with a slim tail and higher peak. For the uniform prior, the lower and upper bounds were taken to be $0$ and $0.2$, respectively. Both the log-normal priors had $\mu = \log 0.05$; the flat-tailed prior had $\sigma = 1$, and the high-peaked prior had $\sigma = 0.1$ as the hyperparameter. The hyperparameter setting for the half-normal prior was the same as in Section 4. We compared performance of the MCPP for the different prior distributions in Scenario 5 and 6 (Table 1): both scenarios considered mixed offspring density ($\alpha_2$=4 and $\alpha_3$=1), one had low bandwidth ($h_2$=0.01 and $h_3$=0.02), and the other had high bandwidth ($h_2$=0.1 and $h_3$=0.01). 

We report the mean absolute percentage bias for estimating the corresponding parameters in the two scenarios for the four different prior settings: i) Half-normal, ii) Uniform, iii) a flat Log-normal, and iv) a tight Log-normal. The half-normal prior-based MCPP model performed the best, and the performance was similar to that for the original model. When the true bandwidth was low, all the models---irrespective of prior choice---generally performed well and similarly to each other, with almost all biases $<8\%$. Differences in performance emerged when the true bandwidth was high, where the analyses with tighter priors produced much less biased estimates ($<10\%$ except in one instance) than the analyses with flatter priors (4-137\%; Table \ref{tab:sense_tab}). However, using an informative log-normal prior backfired even for the low-bandwidth scenario when the offspring density was also low, as for the second offspring process ($\sim$20-25\%). %Further sensitivity analysis regarding the variance parameter for the log-normal prior may be needed to understand its effects.

\begin{table}[ht]
    \centering
    \caption{Results of MCPP based analysis of simulated data, comparing different priors for the bandwidth parameters. The true values for the offspring densities were $\alpha_2$=4 and $\alpha_3$=1. The true values for the bandwidth parameters were $h_2$=0.01 and $h_3$=0.02 under low bandwidth and $h_2$=0.1 and $h_3$=0.01 under high bandwidth. The parent process is denoted $\lambda^C$. Results are presented as mean absolute percentage bias of the estimated parameter values based on posterior means of each of the $100$ simulated datasets.  There were no other taxa unrelated to these multi-layered arrangements.}
    \label{tab:sense_tab}
    \begin{tabular}{rrrrrr}
    \hline
    & & & & Log-normal & Log-normal\\
    & & Half-normal & Uniform & (flat) & (tight)\\
    \hline
    &$\alpha_2$&0.03&0.03&0.03&0.03\\
    Low&$\alpha_3$&0.07&0.07&0.07&0.07\\
    bandwidth&$h_2$&0.02&0.02&0.02&0.05\\
    &$h_3$&0.04&0.04&0.04&0.24\\
    &$\lambda^C_1$&0.06&0.06&0.06&0.06\\
    \hline
    &$\alpha_2$&0.03&0.31&0.52&0.03\\
    High&$\alpha_3$&0.05&0.05&0.05&0.05\\
    bandwidth&$h_2$&0.09&0.99&1.37&0.10\\
    &$h_3$&0.03&0.03&0.03&0.20\\
    &$\lambda^C_1$&0.06&0.06&0.06&0.06\\
    \hline
    \end{tabular}
\end{table}

\clearpage
\section{Additional Figures and Tables for the Analysis of Human Microbiome Biofilm Image Data}

\begin{figure}
    \centering
    \includegraphics[width=\linewidth]{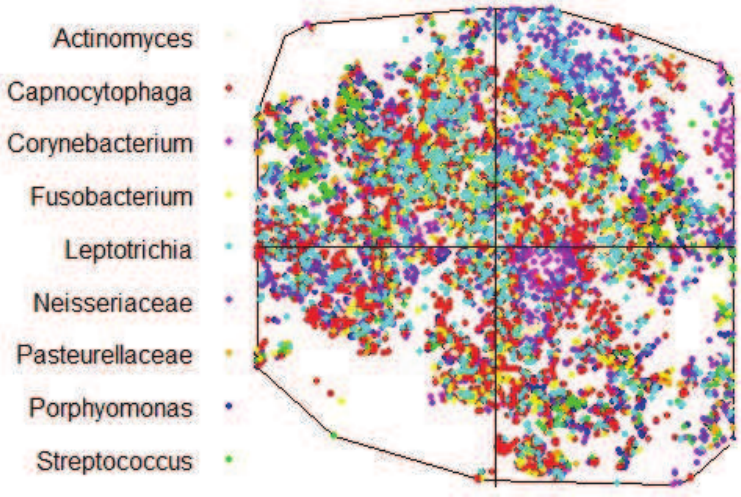}
    \caption{Division of the dental plaque sample image into the first (bottom left), second (top left), third (bottom right) and fourth (top right) quadrants. Black space has been removed.}
    \label{fig: quad}
\end{figure}

\begin{figure}
    \centering
    \includegraphics[width=\linewidth]{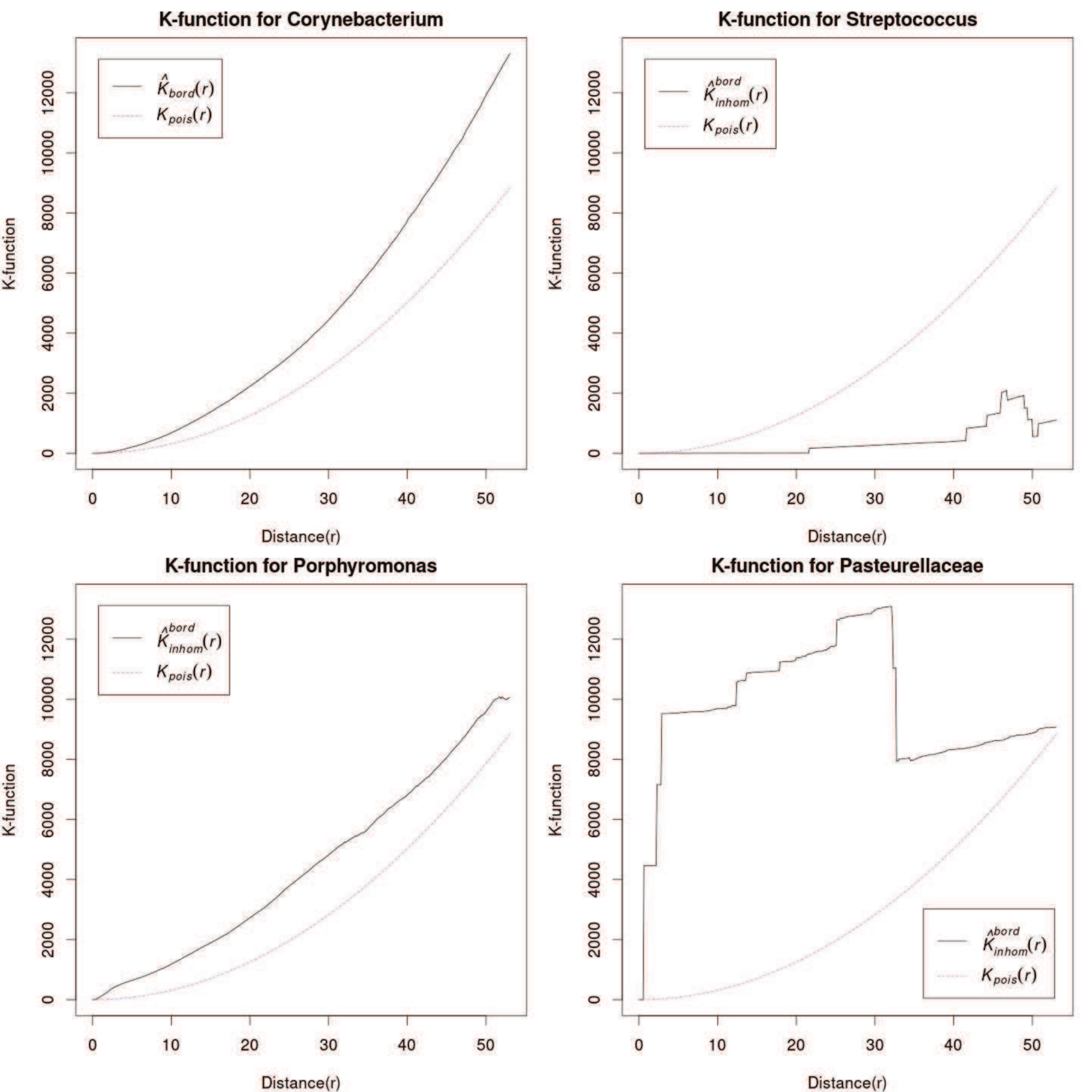}
    \caption{Border-corrected $K$-functions $\big (\hat{K}_{\text{bord}}(r)$ or $\hat{K}_{\text{inhom}}^{\text{bord}}(r)\big)$ for the processes corresponding to \textit{Corynebacterium} (top left), \textit{Streptococcus} (top right), \textit{Porphyromonas} (bottom left) and \textit{Pasteurellaceae} (bottom right) in comparison to that of a homogeneous Poisson process $\big (\hat{K}_{\text{pois}}(r)\big )$.}
    \label{fig:kfuncs_dat}
\end{figure}

\begin{figure}
    \centering
    \includegraphics[width=\linewidth]{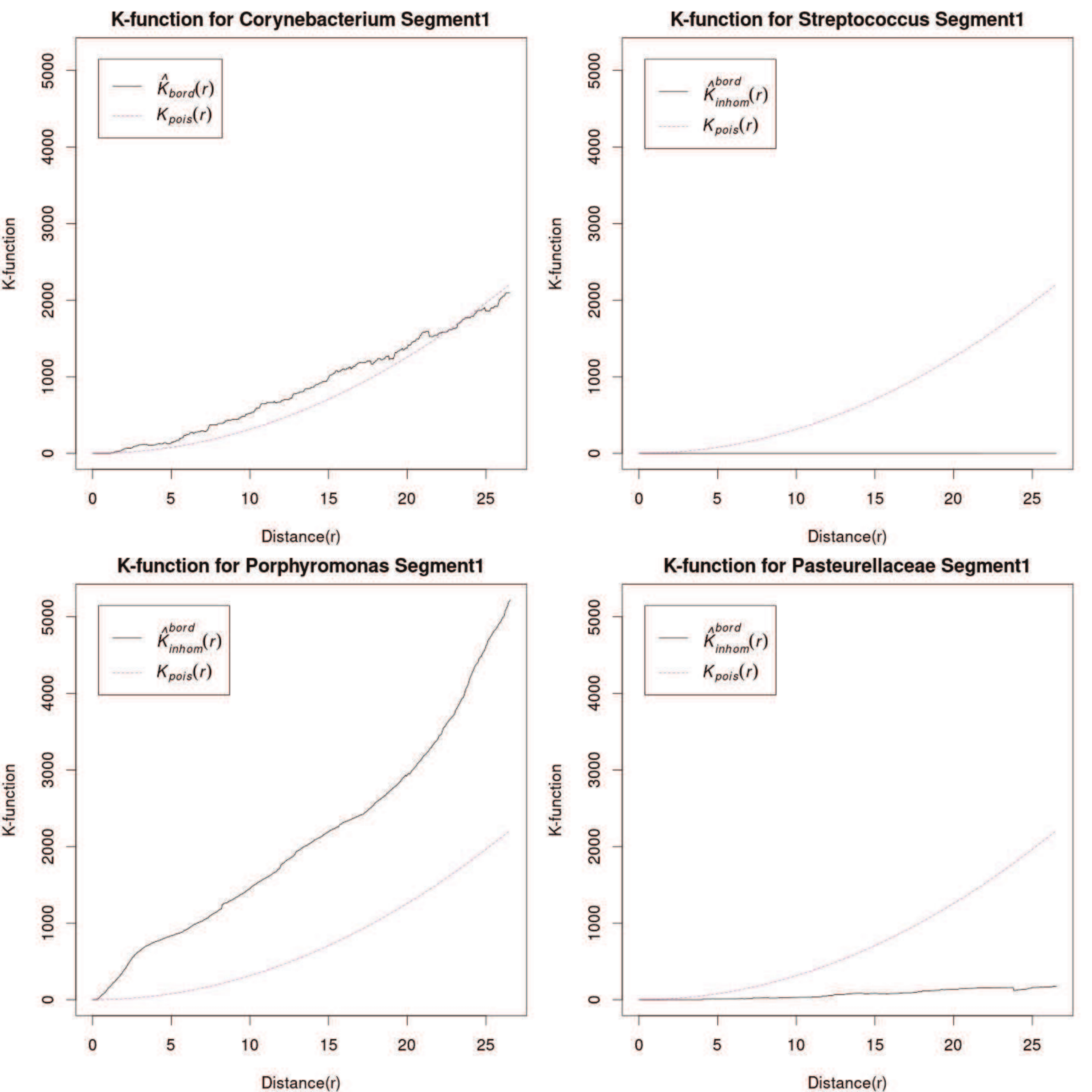}
    \caption{Border-corrected $K$-functions $\big (\hat{K}_{\text{bord}}(r)$ or $\hat{K}_{\text{inhom}}^{\text{bord}}(r)\big)$ for the processes corresponding to \textit{Corynebacterium} (top left), \textit{Streptococcus} (top right), \textit{Porphyromonas} (bottom left) and \textit{Pasteurellaceae} (bottom right) in comparison to that of a homogeneous Poisson process $\big (\hat{K}_{\text{pois}}(r)\big )$ in Segment 1}
    \label{fig:kf_1}
\end{figure}

\begin{figure}
    \centering
    \includegraphics[width=\linewidth]{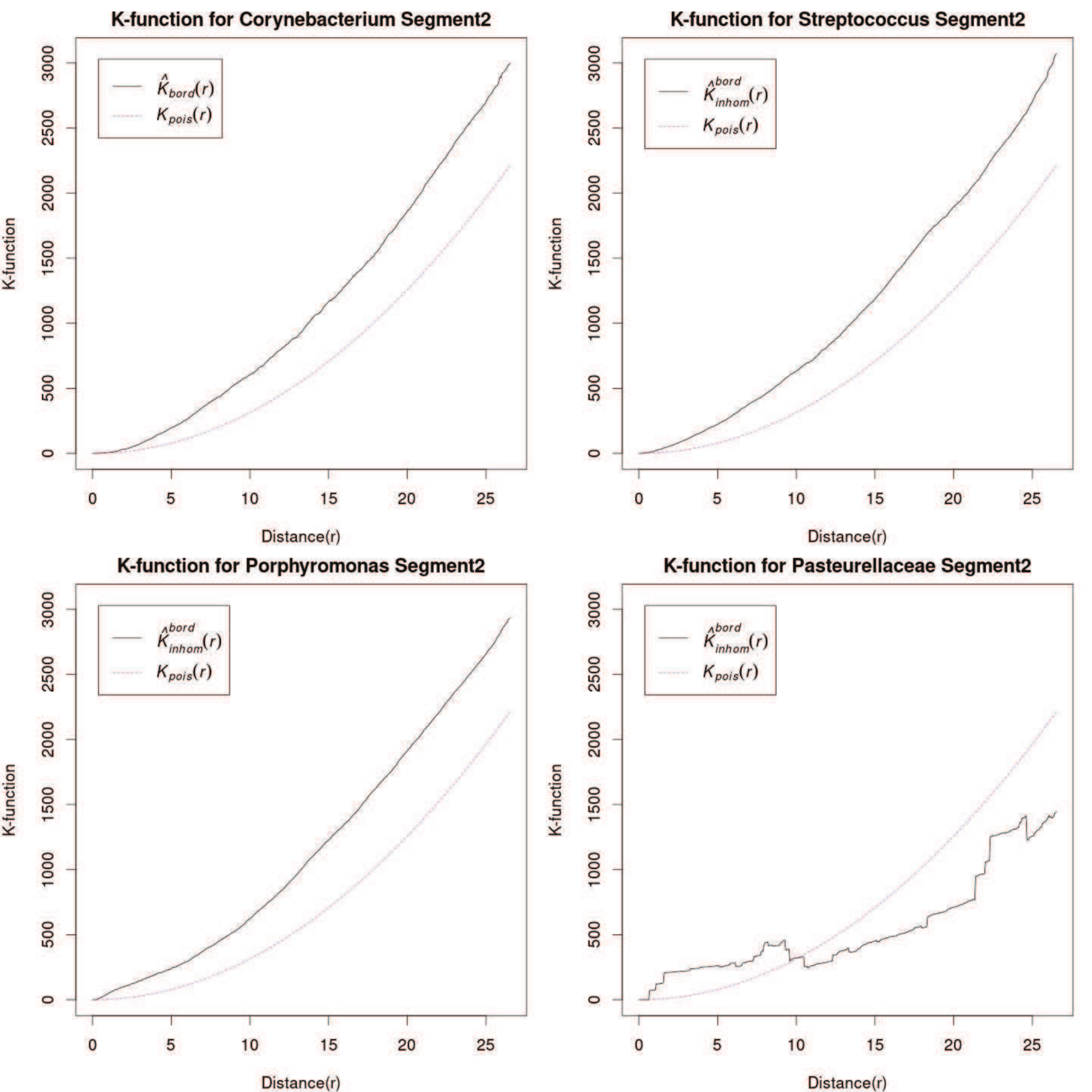}
    \caption{Border-corrected $K$-functions $\big (\hat{K}_{\text{bord}}(r)$ or $\hat{K}_{\text{inhom}}^{\text{bord}}(r)\big)$ for the processes corresponding to \textit{Corynebacterium} (top left), \textit{Streptococcus} (top right), \textit{Porphyromonas} (bottom left) and \textit{Pasteurellaceae} (bottom right) in comparison to that of a homogeneous Poisson process $\big (\hat{K}_{\text{pois}}(r)\big )$ in Segment 2}
    \label{fig:kf_2}
\end{figure}

\begin{figure}
    \centering
    \includegraphics[width=\linewidth]{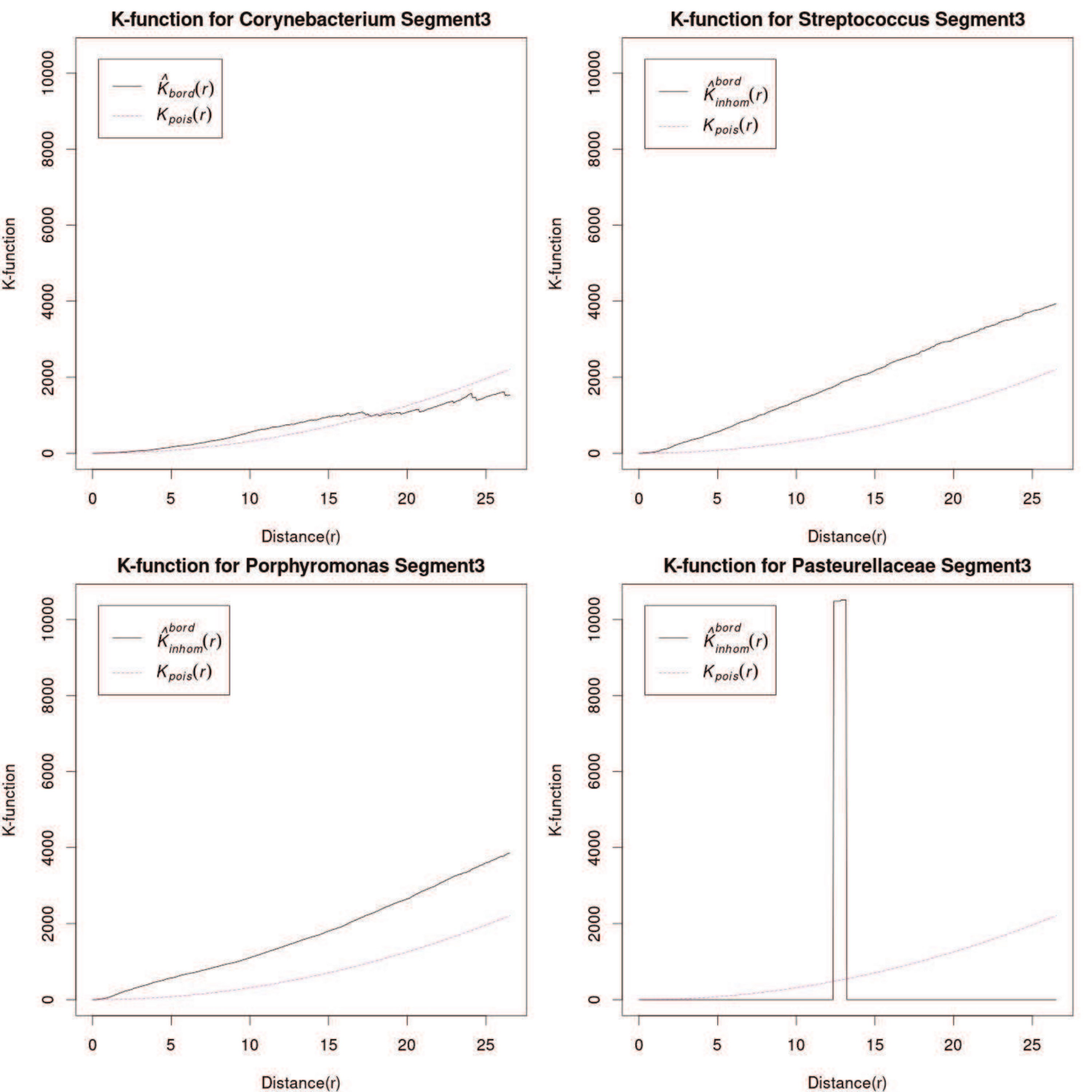}
    \caption{Border-corrected $K$-functions $\big (\hat{K}_{\text{bord}}(r)$ or $\hat{K}_{\text{inhom}}^{\text{bord}}(r)\big)$ for the processes corresponding to \textit{Corynebacterium} (top left), \textit{Streptococcus} (top right), \textit{Porphyromonas} (bottom left) and \textit{Pasteurellaceae} (bottom right) in comparison to that of a homogeneous Poisson process $\big (\hat{K}_{\text{pois}}(r)\big )$ in Segment 3}
    \label{fig:kf_3}
\end{figure}

\begin{figure}
    \centering
    \includegraphics[width=\linewidth]{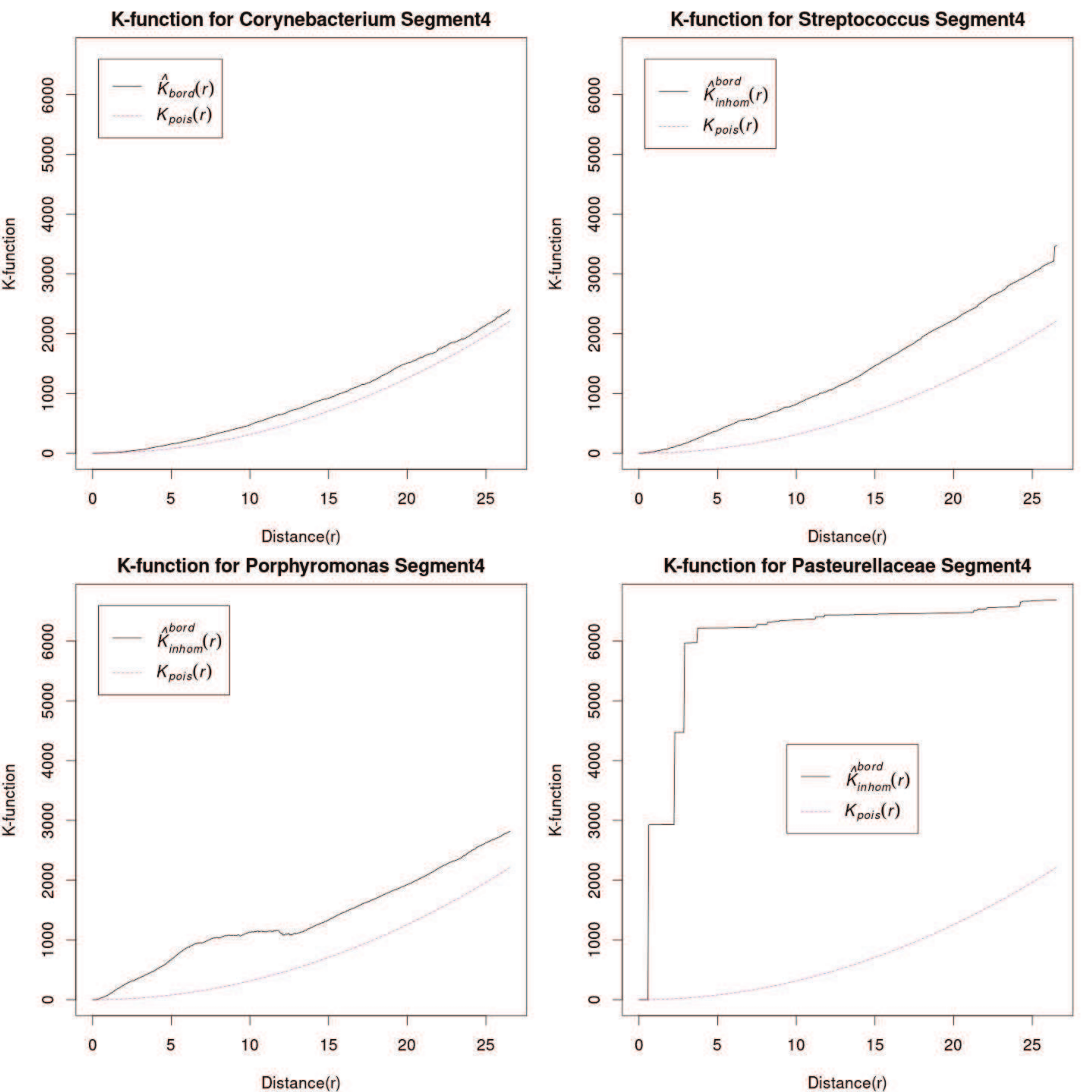}
    \caption{Border-corrected $K$-functions $\big (\hat{K}_{\text{bord}}(r)$ or $\hat{K}_{\text{inhom}}^{\text{bord}}(r)\big)$ for the processes corresponding to \textit{Corynebacterium} (top left), \textit{Streptococcus} (top right), \textit{Porphyromonas} (bottom left) and \textit{Pasteurellaceae} (bottom right) in comparison to that of a homogeneous Poisson process $\big (\hat{K}_{\text{pois}}(r)\big )$ in Segment 4}
    \label{fig:kf_4}
\end{figure}

Here, we present a visual representation of the four quadrants of the whole dental plaque data for subset analyses in Figure \ref{fig: quad} and the abundances of different taxa in the four quadrants of the subsetted data (Table \ref{tab: abundance}). The estimates for the intensity functions for the five taxa (\emph{Neisseriaceae}, \emph{Capnocytophaga}, \emph{Actinomyces}, \emph{Fusobacterium}, \emph{Leptotrichia}) that have no visible spatial relationship with the parent-offspring-type configurations also varied across the quadrants (Table \ref{tab: res_other}). K-functions for the whole and subsetted analyses (Figures \ref{fig:kfuncs_dat} and \ref{fig:kf_1} through \ref{fig:kf_4}) also varied noticeably by quadrant. The DIC estimates for the different models explored in Section 5.2 (Table \ref{tab:data_dic}) indicate that the models with the \emph{Fusobacterium} and \emph{Leptotrichia} as an additional parent-offspring pair is a better fit to the data than the original model, while models fitting \emph{Streptococcus} around \emph{Fusobacterium} do not fit the data well.

\begin{table}[ht]
\centering
%\spacingset{1.175}
\caption{The abundance (counts) of bacterial taxa of interest in the human dental plaque sample image data and its four subdivided quadrants.}
\label{tab: abundance}
\begin{tabular}{rrrrrr}
  \hline
 Taxon & \multicolumn{4}{c}{Quadrant} & Total\\ 
\cline{2-5}
& I & II & III & IV & \\ 
  \hline
\textit{Actinomyces} & 119& 280& 154& 223& 776 \\ 
  \textit{Capnocytophaga} & 512& 755& 574& 573& 2414 \\ 
  \textit{Corynebacterium} & 58& 219& 186& 245& 708 \\  
  \textit{Fusobacterium} & 92& 250& 141& 173& 656 \\ 
  \textit{Leptotrichia} & 191& 411& 234& 339& 1175 \\ 
  \textit{Neisseriaceae} & 339& 479& 402& 491& 1711 \\ 
  \textit{Pasteurellaceae} & 53& 130& 76& 106& 365 \\ 
  \textit{Porphyromonas} & 227& 525& 269& 420& 1441 \\ 
  \textit{Streptococcus} & 98& 379& 163& 249& 889 \\ 
   \hline
\end{tabular}
\end{table}

\begin{table}[h]
\centering
\caption{The posterior means of parameters associated with \textit{Neisseriaceae} ($\lambda_5$), \textit{Capnocytophaga} ($\lambda_6$), \textit{Actinomyces} ($\lambda_7$), \textit{Fusobacterium} ($\lambda_8$) and \textit{Leptotrichia} ($\lambda_9$) obtained by applying the proposed MCPP method on the entire image and on each of the four quadrants of the dental plaque sample image. All results are rounded to two decimal places. The posterior standard deviations were all smaller than 0.01 and are not reported separately.}
\label{tab: res_other}
\begin{tabular}{rrrrrr}
  \hline
 & $\lambda_5$ & $\lambda_6$ & $\lambda_7$ & $\lambda_8$ & $\lambda_9$ \\ 
 \hline
 Full Image & 0.04& 0.06& 0.02 & 0.02& 0.03\\
  \hline
Segment 1 & 0.04 & 0.06 & 0.01 & 0.01 & 0.02\\ 
  Segment 2 & 0.05& 0.07& 0.03 & 0.02 & 0.04\\ 
  Segment 3 & 0.04& 0.05& 0.01& 0.01& 0.02\\ 
  Segment 4 & 0.05& 0.06& 0.02& 0.02& 0.03\\ 
   \hline
\end{tabular}
\end{table}


\begin{table}[]
    \centering
    \caption{The parent-offspring relationships explored in different models and their DIC. The abbreviations C, S, Po, Pa, F and L are used for \emph{Corynebacterium}, \emph{Streptococcus}, \emph{Porphyromonas}, \emph{Pasteurellaceae}, \emph{Fusobacterium} and \emph{Leptotrichia}. The $\rightarrow$ implies parent-offspring relationship with the arrow directed from the parent to the offspring(s).}
    \label{tab:data_dic}
    \begin{tabular}{cccccccc}
    \hline
    Identifier & Parent-Offspring Realtions Present & DIC\\
    \hline
    1 &$C \rightarrow S Po \vdots S \rightarrow Pa$& 124985.4\\
    2 &$C \rightarrow Po \vdots S \rightarrow Pa \vdots F \rightarrow S$& 125531.1\\
    3 &$C \rightarrow S Po \vdots F \rightarrow S \vdots S \rightarrow Pa $& 134007.1\\
    4 &$C \rightarrow S Po \vdots S \rightarrow Pa \vdots F \rightarrow L$& 124317.0\\
    5 &$C \rightarrow S Po \vdots S \rightarrow Pa \vdots L \rightarrow F$& 124303.5\\
    6 &$C \rightarrow S Po \vdots S \rightarrow Pa \vdots F \rightarrow L S$& 133357.2\\
    7 &$C \rightarrow S Po \vdots S \rightarrow Pa \vdots L \rightarrow F \vdots F \rightarrow S$& 133345.2\\
    \hline
    \end{tabular}
\end{table}

\begin{figure}
    \centering
    \includegraphics[width=0.5\linewidth]{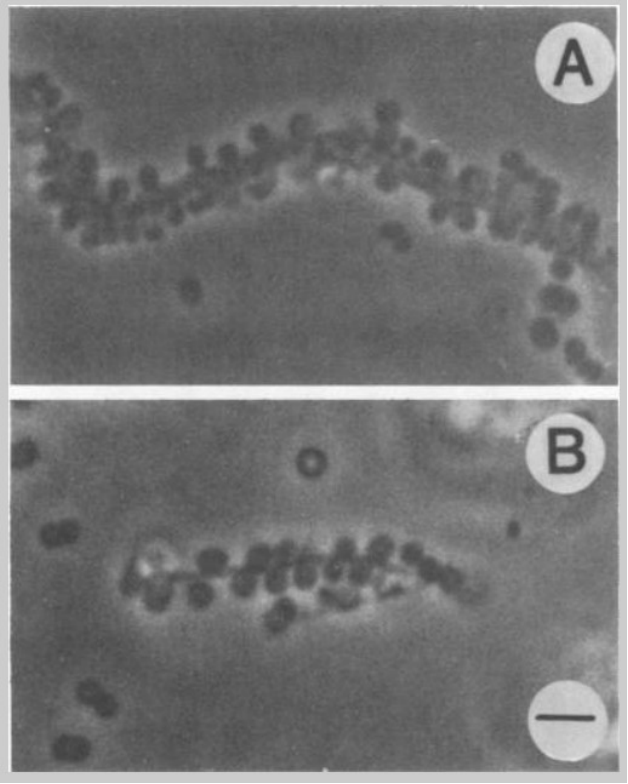}
    \caption{In vitro corncob formation between (A) \emph{S. sanguis CC5A} and \emph{B. matruchotti} and (B) \emph{S. sanguis CC5A} and \emph{F. nucleatum}. The bar is $2 \mu m$. Image taken from \cite{lancy1983corncob}.}
    \label{fig:sus_strep_fuso}
\end{figure}

\begin{figure}
    \centering
    \includegraphics[width=0.5\linewidth]{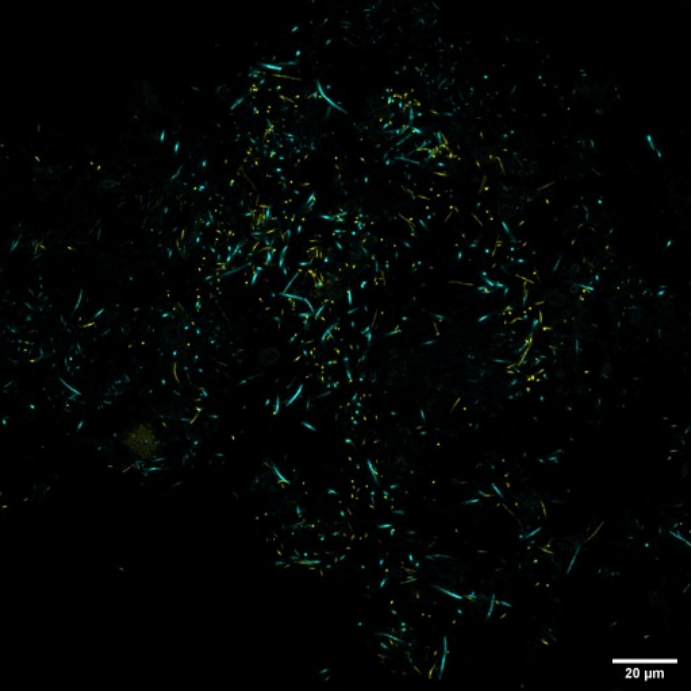}
    \caption{\emph{Fusobacterium} (Yellow) and \emph{Leptotrichia} (Blue) occupying space near each other in the dental plaque sample we analyze.}
    \label{fig:fuso_lepto}
\end{figure}

We additionally present two images (Figures \ref{fig:sus_strep_fuso} and \ref{fig:fuso_lepto}) to make the case for testing the different models as described in Sections 5.2.1 and 5.2.2 of the main manuscript.

\begin{figure}
    \centering
    \begin{subfigure}[b]{0.45\linewidth}
        \caption{Model 1}
        \includegraphics[width=\linewidth]{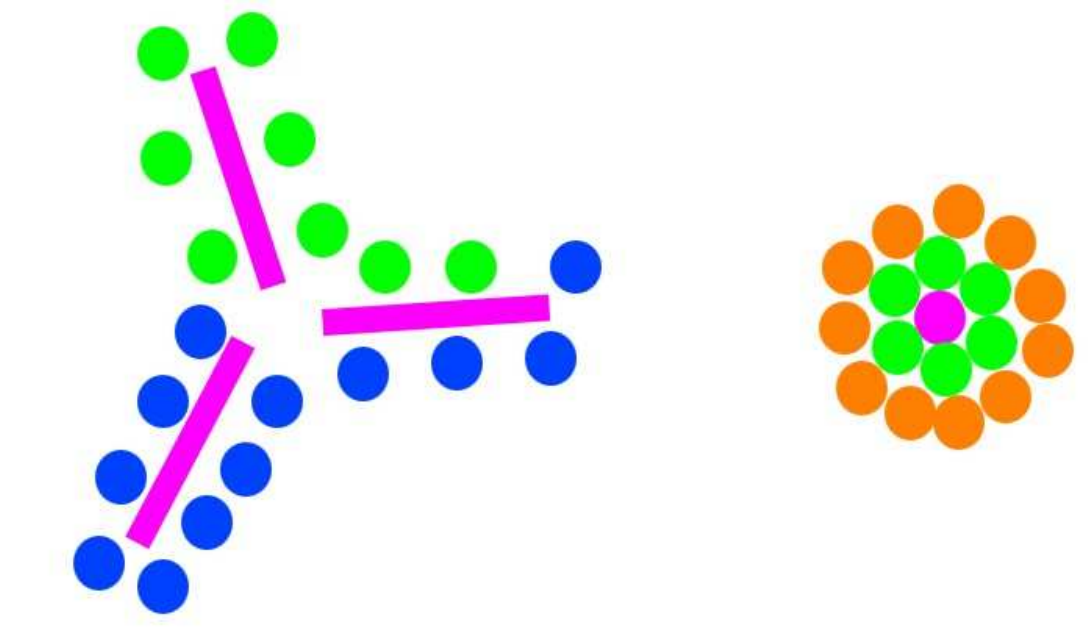}
    \end{subfigure}
    \hfill
    \begin{subfigure}[b]{0.45\linewidth}
        \caption{Model 2}
        \includegraphics[width=\linewidth]{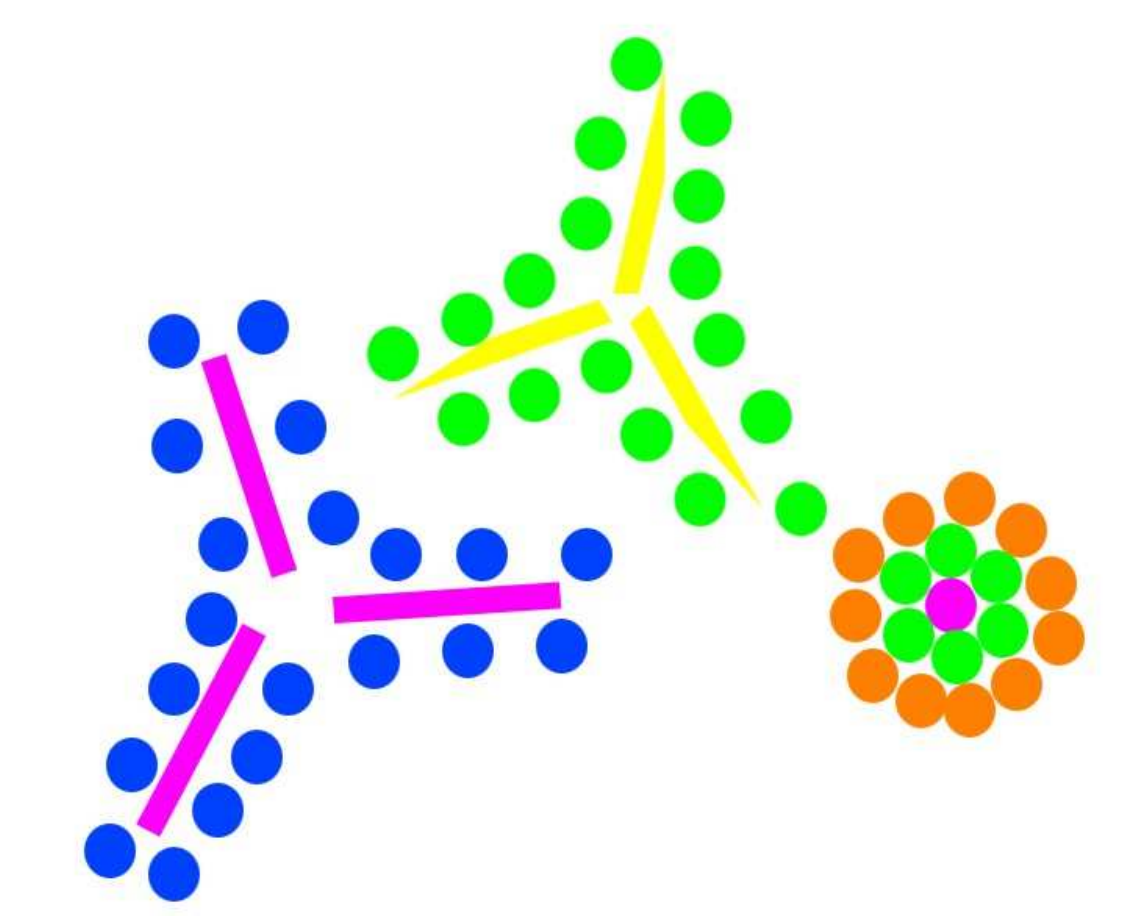}
    \end{subfigure}

    \begin{subfigure}[b]{0.45\linewidth}
    \caption{Model 3}
        \includegraphics[width=\linewidth]{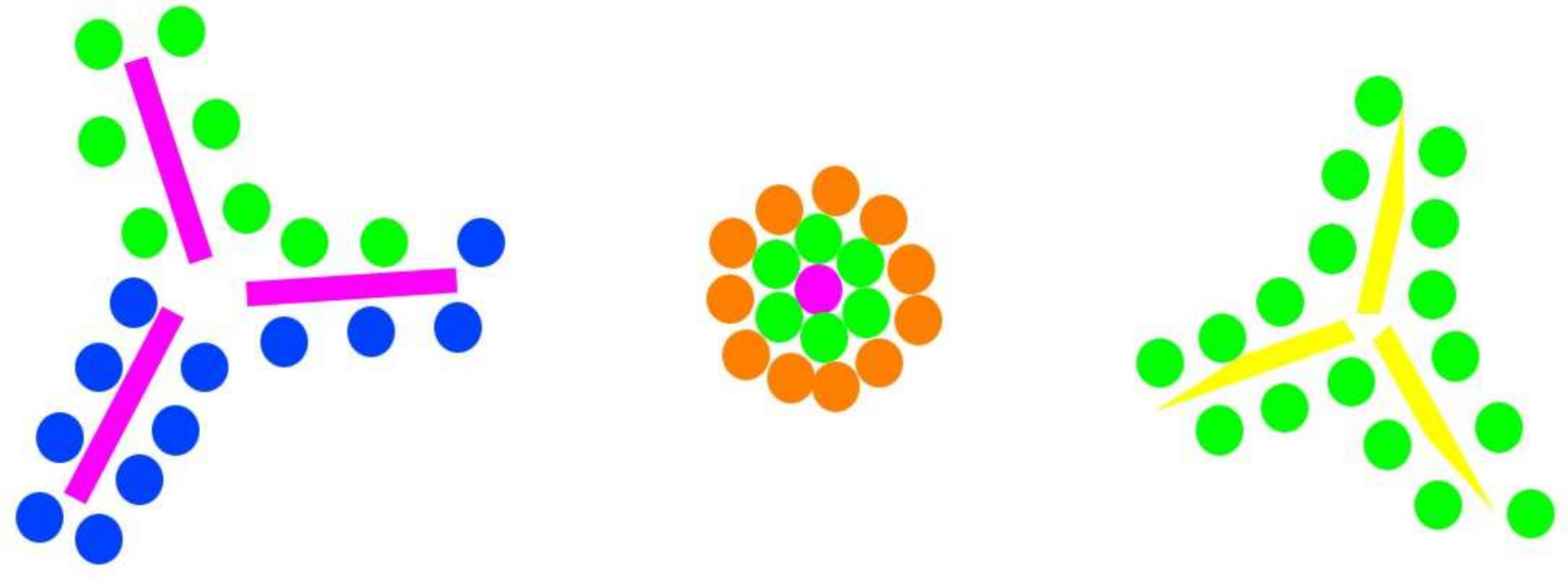}
    \end{subfigure}
    \hfill
    \begin{subfigure}[b]{0.45\linewidth}
        \caption{Model 4}
        \includegraphics[width=\linewidth]{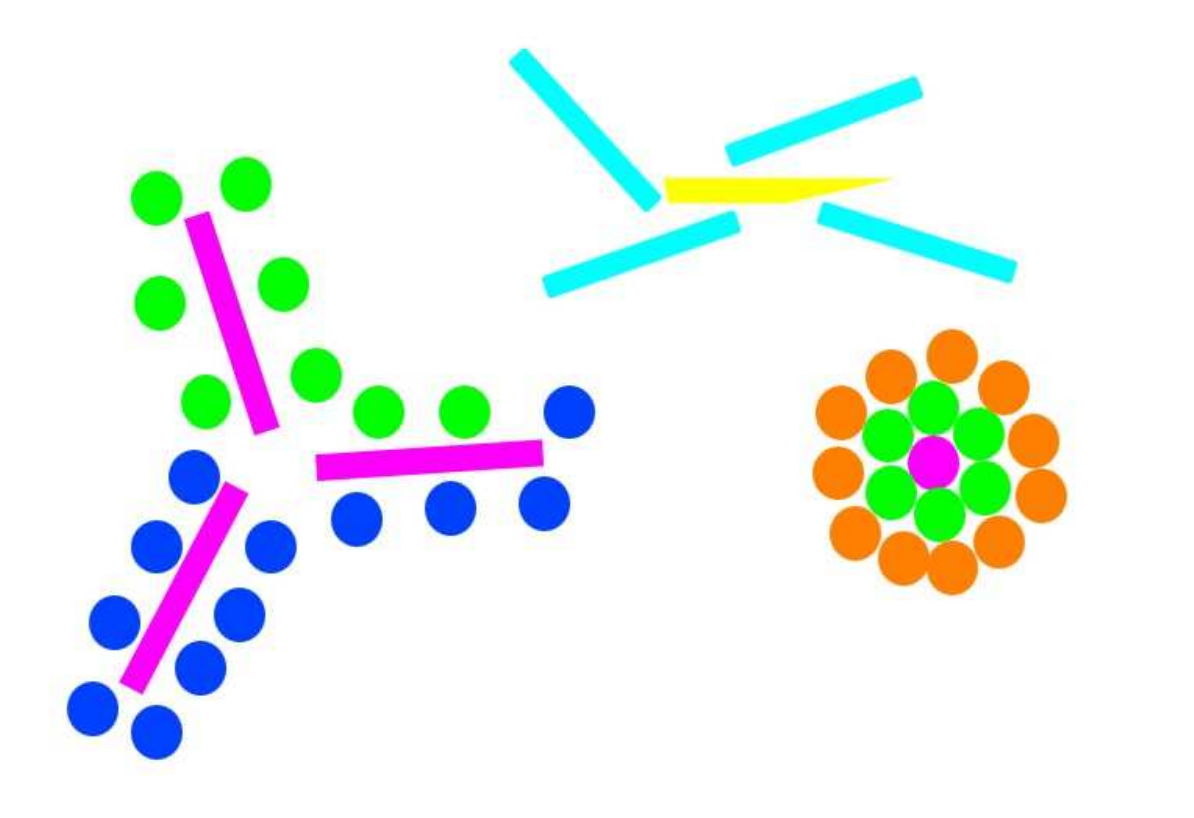}
    \end{subfigure}

    \begin{subfigure}[b]{0.45\linewidth}
        \caption{Model 5}
        \includegraphics[width=\linewidth]{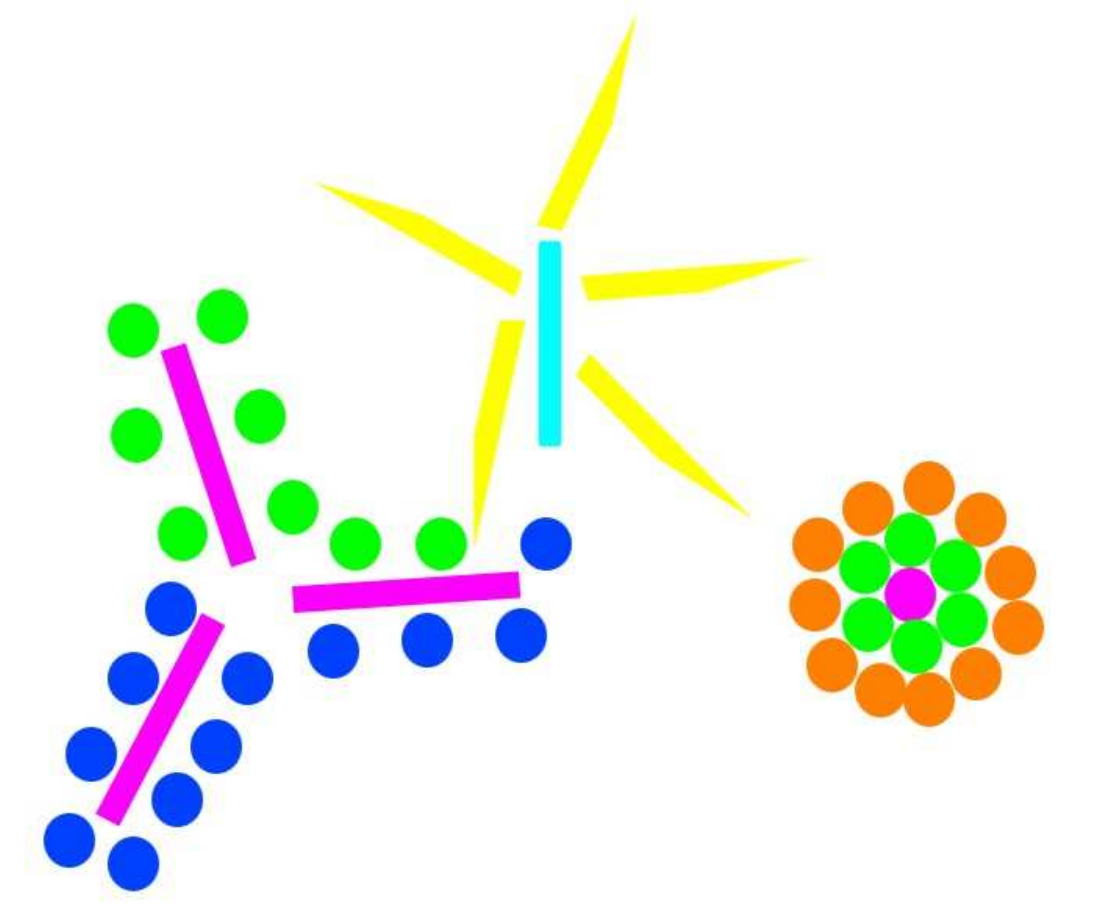}
    \end{subfigure}
    \hfill
    \begin{subfigure}[b]{0.45\linewidth}
        \caption{Model 6}
        \includegraphics[width=\linewidth]{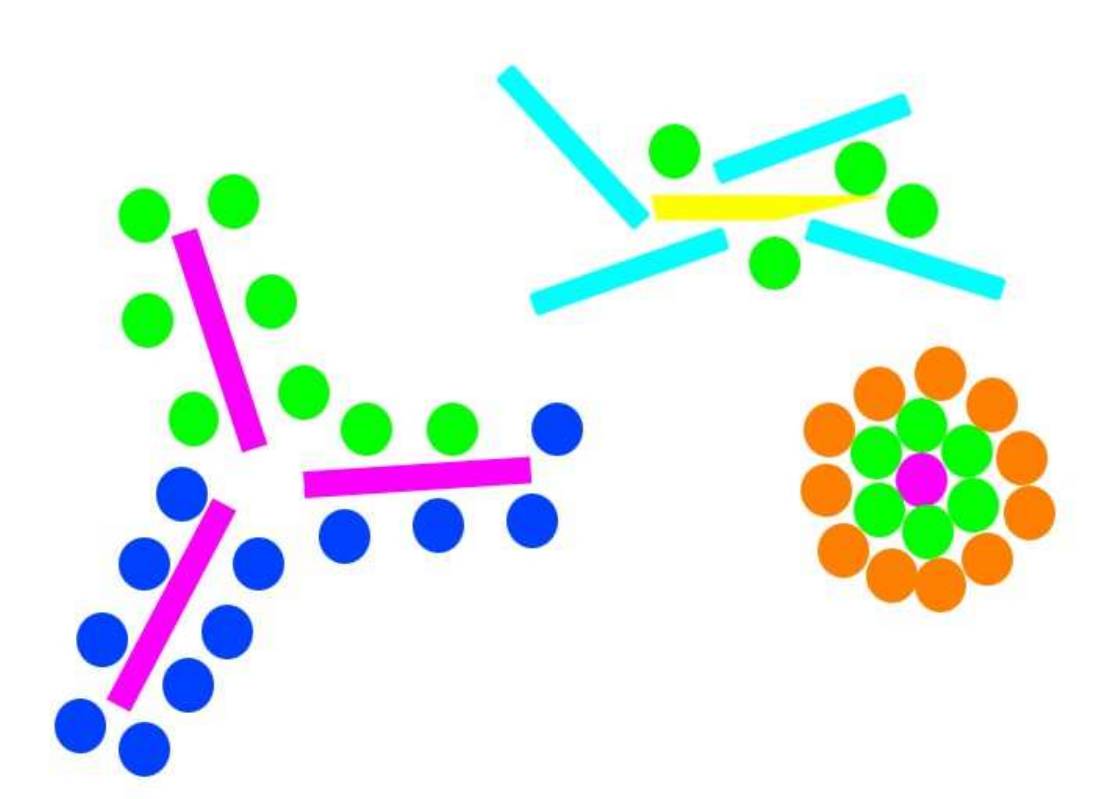}
    \end{subfigure}

    \begin{subfigure}[b]{0.75\linewidth}
        \caption{Model 7}
        \includegraphics[width=\linewidth]{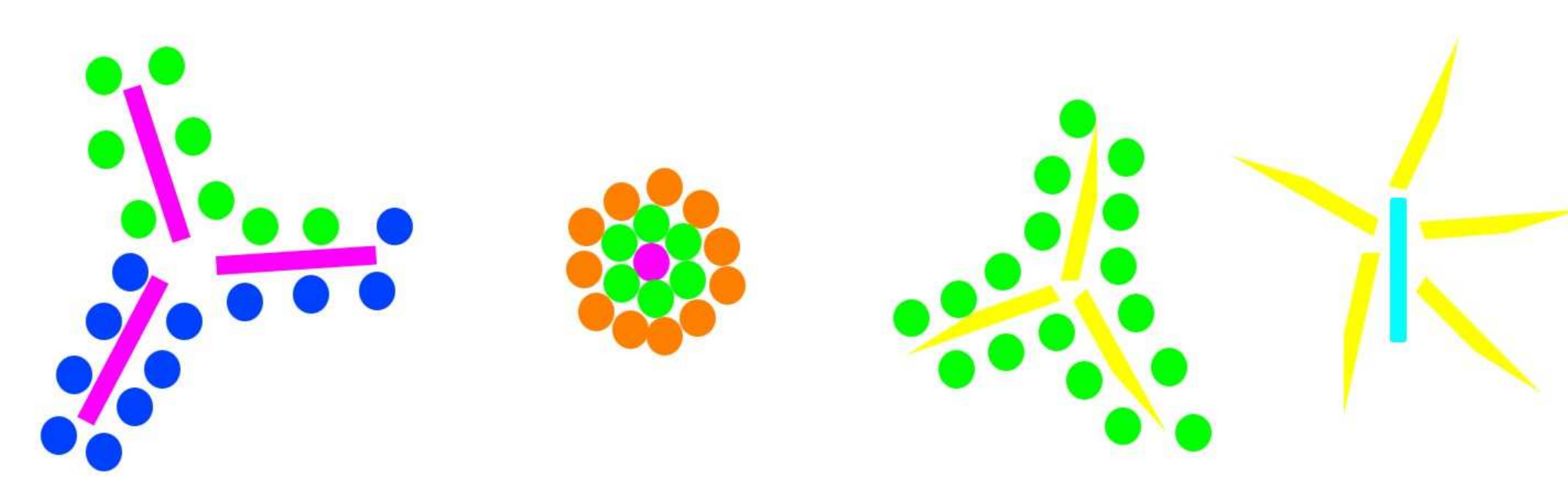}
    \end{subfigure}
    \caption{Cartoon images depicting the different relationships present in the seven different models in Section 5.2.2. These models involve \emph{Corynebacterium} (magenta), \emph{Streptococcus} (green), \emph{Porphyromonas} (blue), \emph{Pasteurellaceae} (Orange), \emph{Fusobacterium} (yellow), and \emph{Leptotrichia} (cyan) in different configurations. Note that for each model, we include here only one illustrative cluster example.}
    \label{fig:enter-label}
\end{figure}

\clearpage

\section{Model Formulation for Given Parent-Offspring Relationships}

In this section, we illustrate the formulation of the MCPP model for a given parent-offspring configuration, as described in Section 3.1 of the main manuscript. Using the notation introduced in Section 3.5, we assume data containing four taxa $A, B, C,$ and $D$ ($m=4$). We detail the construction of the likelihood function according to equations (1)-(3) for the following four different configurations: 
\begin{enumerate}[a)]
    \item $A \rightarrow B \vdots C \rightarrow D$
    \item $A \rightarrow BC$
    \item $A \rightarrow B \vdots B \rightarrow C$
    \item $A \rightarrow BC \vdots C \rightarrow D$
\end{enumerate}

\subsection{Modeling $A \rightarrow B \vdots C \rightarrow D$}
The model represents a configuration involving two parent processes (taxa $A$ and $C$, $p=2$) and two offspring processes (taxa $B$ and $D$, $q=2$). That is, all taxa are interrelated within parent-offspring framework, with no taxa existing outside of these relationships ($m-p-q = 0$). Therefore, following equations (1)-(3), the corresponding model formulation can be written as follows: 
\beq
\begin{split}
    \lambda_A(\bs) &= \lambda_A\\
    \lambda_C(\bs) &= \lambda_B\\
    \lambda_B(\bs) &= \alpha_B\sum_{\bc \in A} k_B(\bs - \bc,h_B)\\
    \lambda_D(\bs) &= \alpha_D\sum_{\bc \in C} k_D(\bs - \bc,h_D). \nonumber
\end{split}
\eeq  
It follows that
\beq
\begin{split}
    l(Y|\btheta) 
    &\propto |\mathcal{W}| - |\mathcal{W}|\lambda_A - |\mathcal{W}|\lambda_C \\ 
    &\quad + \alpha_B\sum_{\bc \in A} \int_{\mathcal{W}} k_B(\bu - \bc,h_B) \,\ d\bu + \alpha_D\sum_{\bc \in C} \int_{\mathcal{W}} k_D(\bu - \bc,h_D) \,\ d\bu\\
    &\quad + n_A\log \lambda_A + n_C \log \lambda_C \\
    &\quad + \sum_{\bs \in B} \log \left( \alpha_B\sum_{\bc \in A} k_B(\bs - \bc,h_B) \right) + \sum_{\bs \in D} \log \left( \alpha_D\sum_{\bc \in C} k_D(\bs - \bc,h_D) \right). \nonumber
\end{split}
\eeq

\subsection{Modeling $A \rightarrow BC$}

In this model, we investigate the configuration where two offspring processes (taxa $B$ and $C$, $q=2$) share the same parent process (taxon $A$, $p=1$). Additionally, the model includes a separate process for taxon $D$, which remains uninvolved in any parent-offspring relationship ($m-p-q=1$). The model formulation can be written as 
\beq
\begin{split}
    \lambda_A(\bs) &= \lambda_A,\\
    \lambda_B(\bs) &= \alpha_B\sum_{\bc \in A} k_B(\bs - \bc,h_B),\\
    \lambda_C(\bs) &= \alpha_C\sum_{\bc \in A} k_C(\bs - \bc,h_C),\\
    \lambda_D(\bs) &= \lambda_D. \nonumber
\end{split}
\eeq 
It follows that
\beq
\begin{split}
    l(Y|\btheta) 
    &\propto |\mathcal{W}| - |\mathcal{W}|\lambda_A - |\mathcal{W}|\lambda_D\\
    &\quad - \alpha_B\sum_{\bc \in A} \int_{\mathcal{W}}k_B(\bu - \bc,h_B) \,\ d\bu - \alpha_C\sum_{\bc \in A} \int_{\mathcal{W}}k_C(\bu - \bc,h_C) \,\ d\bu \\
    &\quad + n_A\log \lambda_A + n_D \log \lambda_D\\
    &\quad + \sum_{\bs \in B} \log \left( \alpha_B\sum_{\bc \in A} k_B(\bs - \bc,h_B) \right) + \sum_{\bs \in C} \log \left( \alpha_C\sum_{\bc \in A} k_C(\bs - \bc,h_C) \right). \nonumber
\end{split}
\eeq

\subsection{Modeling $A \rightarrow B \vdots B \rightarrow C$}
\label{sec:conf3}
This scenario presents a more complex relationship where taxon $B$ behaves both as an offspring to taxon $A$ and as a parent to taxon $C$. In this setup, there is only one process serving as a parent (taxon $A$, $p=1$), two processes functioning as offspring (taxa $B$ and $C$, $q=2$), and a separate process for taxon $D$, which is uninvolved in any parent-offspring relationship ($m-p-q=1$). It is worth noting that $p=1$ as the process for taxon $B$ is counted as an offspring process, while the locations of taxon $B$ will be used for modeling the process for taxon $C$. Therefore, the MCPP under the configuration can be written as 
\beq
\begin{split}
    \lambda_A(\bs) &= \lambda_A,\\
    \lambda_B(\bs) &= \alpha_B\sum_{\bc \in A} k_B(\bs - \bc,h_B),\\
    \lambda_C(\bs) &= \alpha_C\sum_{\bc \in B} k_C(\bs - \bc,h_C),\\
    \lambda_D(\bs) &= \lambda_D. \nonumber
\end{split}
\eeq 
It follows that
\beq
\begin{split}
    l(Y|\btheta) 
    &\propto |\mathcal{W}| - |\mathcal{W}|\lambda_A - |\mathcal{W}|\lambda_D \\
    &\quad -\alpha_B\int_{\mathcal{W}}\sum_{\bc \in A} k_B(\bu - \bc,h_B) \,\ d\bu - \alpha_C\int_{\mathcal{W}}\sum_{\bc \in A} k_C(\bu - \bc,h_C) \,\ d\bu \\
    &\quad + n_A\log \lambda_A  + n_D\log \lambda_D \\
    &\quad +\sum_{\bs \in B} \log \left( \alpha_B\sum_{\bc \in A} k_B(\bs - \bc,h_B) \right) +\sum_{\bs \in C} \log \left( \alpha_C\sum_{\bc \in A} k_C(\bs - \bc,h_C) \right). \nonumber
\end{split}
\eeq

\subsection{Modeling $A \rightarrow BC \vdots C \rightarrow D$}

In this model, we explore a more complex scenario where the two processes for taxa $B$ and $C$ share the same parent process (for taxon $A$). Furthermore, the process for taxon $C$ not only functions as an offspring process with respect to that of taxon $A$ but also acts as a parent process for taxon $D$. This configuration involves one process serving exclusively as a parent (taxon $A$, $p=1$) and three processes serving as offspring processes (taxa $B, C$ and $D$, $q=3$). Consequently, no taxon remains uninvolved in parent-offspring relationships ($m-p-q=0$). As in Section \ref{sec:conf3}, it is important to note that $p=1$ as the process for taxon $C$ is counted as an offspring process, despite its role as a parent process for taxon $D$. The corresponding model formulation can be written as follows:
\beq
\begin{split}
    \lambda_A(\bs) &= \lambda_A,\\
    \lambda_B(\bs) &= \alpha_B\sum_{\bc \in A} k_B(\bs - \bc,h_B),\\
    \lambda_C(\bs) &= \alpha_C\sum_{\bc \in A} k_C(\bs - \bc,h_C),\\
    \lambda_D(\bs) &= \alpha_D\sum_{\bc \in C} k_D(\bs - \bc,h_D).\nonumber
\end{split}
\eeq 
It follows that
\beq
\begin{split}
    l(Y|\btheta) 
    &\propto |\mathcal{W}| - |\mathcal{W}|\lambda_A \\
    &\quad - \alpha_B\int_{\mathcal{W}}\sum_{\bc \in A} k_B(\bu - \bc,h_B) \,\ d\bu - \alpha_C\int_{\mathcal{W}}\sum_{\bc \in A} k_C(\bu - \bc,h_C) \,\ d\bu\\
    &\quad - \alpha_D\int_{\mathcal{W}}\sum_{\bc \in C} k_D(\bu - \bc,h_D) \,\ d\bu + n_A\log \lambda_A + \sum_{\bs \in B}\log \left( \alpha_B\sum_{\bc \in A} k_B(\bs - \bc,h_B)\right)\\
    &\quad + \sum_{\bs \in C}\log \left( \alpha_C\sum_{\bc \in A} k_C(\bs - \bc,h_C)\right) + \sum_{\bs \in D}\log \left( \alpha_D\sum_{\bc \in C} k_D(\bs - \bc,h_D)\right). \nonumber
\end{split}
\eeq

%\backmatter
\clearpage
%\section*{References}
\bibliographystyle{plainnat}
\bibliography{WileyNJD-AMA}

%\appendix

%\backmatter

%\label{lastpage}